\documentclass[a4paper,12pt]{amsart}
\usepackage[foot]{amsaddr}

\theoremstyle{definition}

\theoremstyle{remark}

\usepackage[margin=2.7cm]{geometry}
\usepackage{systeme}
\usepackage{caption}
\usepackage{subcaption}
\usepackage{graphicx}
\usepackage[]{units}
\usepackage{color}
\usepackage{mathrsfs}
\usepackage[ansinew]{inputenc}
\usepackage{float}
\usepackage{lineno}
\usepackage{bm}  
\usepackage{amsfonts} 
\usepackage{esint}
\usepackage{amsmath}
\usepackage{amssymb}
\usepackage{framed} 
\usepackage{multicol} 
\usepackage{tabularx}
\usepackage{tabulary}
\usepackage{amsfonts}
\usepackage{booktabs}
\usepackage{siunitx}
\usepackage{caption}
\usepackage{float}
\usepackage{url} 
\usepackage{bm}
\usepackage{nomencl} 
\usepackage{breakurl} 
\usepackage[breaklinks]{hyperref} 
\usepackage{algorithm,algorithmic}
\usepackage[final]{changes}

\begin{document}
\renewcommand\texteuro{-} 
\nomenclature{$\text{ROM}$}{Reduced Order model}
\nomenclature{$\text{POD}$}{Proper Orthogonal Decomposition}


\title[Coupling of a reduced order CFD model with system code RELAP5]{Development of a coupling between a system thermal-hydraulic code and a reduced order CFD model}

\author{S. Kelbij Star\textsuperscript{1,2*}}
\address{\textsuperscript{1}SCK$\cdot$CEN, Institute for Advanced Nuclear Systems, Boeretang 200, 2400 Mol, Belgium.}
\address{\textsuperscript{2}Ghent University, Department of Electromechanical, Systems and Metal Engineering, Sint-Pietersnieuwstraat 41, B-9000 Ghent, Belgium}
\thanks{\textsuperscript{*}Corresponding Author.}
\email{kelbij.star@sckcen.be}

\author{Giuseppe Spina\textsuperscript{1,3}}
\address{\textsuperscript{3}Department of Civil and Industrial Engineering - University of Pisa - Largo Lucio Lazzarino 2, 56122 Pisa, Italy.}
\email{g.spina11@studenti.unipi.it}

\author{Francesco Belloni\textsuperscript{1}}
\email{francesco.belloni@sckcen.be}


\author{Joris Degroote\textsuperscript{2}}
\email{Joris.Degroote@UGent.be}


\date{}

\dedicatory{}

\begin{abstract}
	The nuclear community has coupled several three-dimensional Computational Fluid Dynamics (CFD) solvers with one-dimensional system thermal-hydraulic (STH) codes. This work proposes to replace the CFD solver by a reduced order model (ROM) to reduce the computational cost. The system code RELAP5-MOD3.3 and a ROM of the finite volume CFD solver OpenFOAM are coupled by a partitioned domain decomposition coupling algorithm using an implicit coupling scheme. The velocity transported over a coupling boundary interface is imposed in the ROM using a penalty method. The coupled models are evaluated on open and closed pipe flow configurations. The results of the coupled simulations with the ROM are close to those with the CFD solver. Also for new parameter sets, the coupled RELAP5/ROM models are capable of predicting the coupled RELAP5/CFD results with good accuracy. Finally, coupling with the ROM is 3-5 times faster than coupling with the CFD solver.
\end{abstract}


\maketitle

\section{Introduction}\label{sec:intro}
The Generation IV innovative nuclear systems cooled by heavy liquid metal are the subject of an ongoing interest demonstrated by a large number projects in progress. One of them is MYRRHA, an experimental fast-spectrum irradiation facility featuring a pool-type primary cooling system operating with molten Lead-Bismuth Eutectic (LBE), that is currently being developed by SCK$\cdot$CEN, a nuclear research institution in Belgium~\cite{abderrahim2012multi}. 

For the design and safety assessment of a new generation of nuclear reactors, computer codes have been developed for the thermal-hydraulic analyses of the reactor's primary system in operational and accident conditions. There are two main types of numerical codes for thermal-hydraulic analyses used in the nuclear industry: the system codes, also called the lumped parameter codes, based on one-dimensional (1D) models of physical transport phenomena and the field codes, based on three-dimensional (3D) Computational Fluid Dynamics (CFD) models~\cite{OECD}. The system codes are, in general, based upon the solution of six balance equations for liquid and vapor. In addition, they use (quasi-steady state) heat transfer correlations to model the heat transfer between a solid, such as tubes or structures, and its surrounding fluid~\cite{petruzzi2008thermal,bertolotto2009single}. 

The flow in many reactor primary components exhibit phenomena as natural circulation, mixing and stratification that cannot be modeled by system codes adequately. CFD codes are therefore used to numerically simulate these types of transient flows to accurately quantify the system behavior in accident conditions and to handle complex geometries~\cite{toti2018coupled}. However, the number of nuclear reactor simulations in a safety analysis is, in the majority of cases, beyond the possibilities of present hardware if a CFD code is used alone.

Thus, to get the best out of both worlds, coupling between system and CFD codes has been postulated as a new method for thermal-hydraulic analyses. The nuclear community has performed extensive research on interfacing CFD codes with the traditional system codes. Gibling and Mahaffy~\cite{gibeling2002benchmarking} were among the first to study the transition between the 1D and 3D descriptions at an interface. 

A well recognized system thermal-hydraulic (STH) code by many nuclear authorities for safety analyses is the RELAP5 series. The RELAP5 series have been coupled with several computer codes as the sub-channel code  COBRA-TF~\cite{lee1992cobra,jeong1997development}, the containment analysis code GOTHIC~\cite{grgic2004coupled,huang2016performance} and CFD codes ANSYS-CFX (previous called CFDS-FLOW3D)~\cite{burns1988harwell,aumiller2001coupled}, ANSYS Fluent~\cite{schultz2003coupling,FENG2017,li2014preliminary, anderson2008analysis, angelucci2017sth} and Star-CCM+~\cite{jeltsov2013development}. Recently, work has been conducted in the framework of the THINS project of the 7th Framework EU Program on nuclear fission safety~\cite{bandini2015assessment,PIALLA2015}. 

SCK$\cdot$CEN uses the RELAP5-3D~\cite{relap53d} version for MYRRHA safety studies that allows the use of LBE as a working fluid. Moreover, SCK$\cdot$CEN has developed a numerical algorithm to couple RELAP5-3D with ANSYS Fluent for multi-scale transient simulations of pool-type reactors~\cite{toti2018coupled}. Another CFD code that has been coupled already to several STH codes~\cite{PIALLA2015,fiorina2015gen,zhang2020multiscale}, but to the best of the authors' knowledge not yet with RELAP5, is the open source code OpenFOAM (OF)~\cite{Jasak}. 

Even though coupled systems require considerably less computational resources and time than stand-alone CFD codes, the gain in computational effort is still limited by the CFD part~\cite{bury2017coupling}. To overcome this burden, this work proposes to couple the system code with a reduced order model (ROM) of the high fidelity CFD code. 

The basic idea of reduced order CFD modeling is to retain the essential physics and dynamics of a high fidelity CFD model by projecting the (discretized) equations describing the fluid problem onto a low-dimensional basis~\cite{hesthaven2016certified, quarteroni2015reduced,veroy2003reduced,rozza2007reduced}. This basis contains only the essential features of a number of solutions of the high-fidelity simulations. Therefore, the reduced order model contains a lower number of degrees of freedom than the high fidelity models. That way, they are computationally more efficient, but have generally a lower accuracy than the high fidelity models~\cite{Lassila,grepl2007efficient}. Parametric ROMs can be used for evaluating solutions on new sets of parameter values or for time evolution that are different from those of the original simulations~\cite{benner2015survey,gunzburger2007reduced,fick2017reduced}. Therefore, reduced order models are suitable for control purposes or sensitivity analyses that require results of a large number of simulations for different parameter values.

In this work, the STH code RELAP5-MOD3.3~\cite{relap5mod} and a reduced order CFD model that is constructed using the libraries of the open source code OpenFOAM 6 are coupled, which is called the RELAP5/ROM model hereafter. The codes are coupled using a partitioned domain decomposition coupling algorithm, which is explained in Section~\ref{sec:coupling}. The exchange of the hydraulic quantities between the coupled domains at the coupling interfaces is explained in Section~\ref{sec:interfaces}. The CFD and ROM formulations for an incompressible Newtonian fluid are described in Sections~\ref{sec:codes} and~\ref{sec:ROM}, respectively. In Section ~\ref{sec:challenges}, some challenges for coupling STH codes with reduced order models are presented. The set-up of three numerical test cases, the open pipe flow test, the open pipe flow reversal test and the closed pipe flow test, are described in Section~\ref{sec:setup}. Then in Section~\ref{sec:results}, the coupling methodology is first evaluated by comparing the results of a coupled RELAP5/CFD model with RELAP5 stand-alone results. Consecutively, the coupled RELAP5/ROM model is tested on a series of parametric problems that are evaluated against the coupled RELAP5/CFD model and the results are discussed in Section~\ref{sec:discussion}. Finally, conclusions are drawn and an outlook for further improvements is provided in Section~\ref{sec:conclusion}.

\section{Coupling methodology}\label{sec:coupling}
A methodology is developed for a partitioned coupling approach~\cite{matthies2003strong} together with a domain decomposition method~\cite{smith2004domain} in which the different domains are resolved separately by independent solvers. The whole simulation domain is split into sub-domains; where the one-dimensional approximation is deemed accurate enough for the given problem, the sub-domain is allocated to the STH code and if not to the CFD code. The number of coupling faces between the sub-domains is identified. 

At each coupling interface between two sub-domains, thermal-hydraulic quantities, like mass flow rate, pressure and temperature, are exchanged between the solvers. By treating the sub-domains as black boxes, the following input-output relations hold at each coupling interface
\begin{equation}\label{eq:inout1}
\boldsymbol{O}_{STH} = G_{STH}(\boldsymbol{I}_{STH}),
\end{equation}
\begin{equation}\label{eq:inout2}
\boldsymbol{O}_{CFD} = G_{CFD}(\boldsymbol{I}_{CFD}),
\end{equation}
\noindent where $\boldsymbol{I}$ and $\boldsymbol{O}$ are the input and output vectors, respectively. These vectors are either obtained by the STH code or the CFD code. $G$ is the associated operator. Hence, the output $\boldsymbol{O}$ is obtained from the given input $\boldsymbol{I}$.

Then, as the thermal-hydraulic quantities are exchanged between the sub-domains, the following relation holds between the inputs and outputs at the coupling interfaces
\begin{equation}\label{eq:interface1}
\boldsymbol{I}_{CFD} = \boldsymbol{O}_{STH},
\end{equation}
\begin{equation}\label{eq:interface2}
\boldsymbol{I}_{STH} = \boldsymbol{O}_{CFD}.
\end{equation}
Based on Equations~\ref{eq:inout1}-\ref{eq:interface2}, the STH/CFD coupled problem can be expressed in its fixed-point formulation as follows
\begin{equation}\label{eq:fixedpoint}
\boldsymbol{I}_{CFD} = G_{STH}(\boldsymbol{I}_{STH}) = G_{STH}(G_{CFD}(\boldsymbol{I}_{CFD})).
\end{equation}
For time-dependent problems, the coupling can either be done by an explicit or implicit coupling method. Explicit coupling procedures are appealing in terms of efficiency as only one (or a few) solution of the sub-problems per time step are needed. However, the numerical stability of the scheme can be drastically compromised. This especially applies to incompressible fluids~\cite{toti2016development} for which even a small change in pressure drop will immediately affect the whole solution domain as the fluid density does not change with pressure unlike in compressible fluids. This phenomenon is well-known from coupling schemes in fluid-structure interaction problems~\cite{farhat2010robust,van2011partitioned}. Consequently, the time step size needs to be restricted. For more details and the analytical explanation, the reader is referred to~\cite{grunloh2016novel}. Moreover, it is known from the work of Toti et al.~\cite{toti2016development} that the implicit coupling scheme is numerically more stable than the explicit coupling schemes. As reduced order models are sensitive to numerical instabilities~\cite{Akhtar,Sirisup,bergmann2009enablers}, the coupling is only done by an implicit coupling method in this work.

\subsection{Implicit coupling numerical scheme}
In order to assure a global conservation of transported quantities over the interface, an implicit numerical scheme, which determines the solution of the fixed-point problem of Equation~\ref{eq:fixedpoint}, is implemented. The exchange of data between the codes is repeated through an iterative procedure within a time step until a defined convergence criterion is met. In this way, an equilibrium is reached at the coupling boundary interfaces and the numerical stability is improved. 
The coupled problem of Equation~\ref{eq:fixedpoint} is reformulated as a root finding problem:
\begin{equation}\label{eq:implicit_R}
\boldsymbol{R}(\boldsymbol{I}_{CFD}) =  G_{STH}(G_{CFD}(\boldsymbol{I}_{CFD})) - \boldsymbol{I}_{CFD} = \boldsymbol{0},
\end{equation}
where $\boldsymbol{R}$ is the residual vector. This residual is approximated by a first order Taylor expansion around the current solution at each coupling iteration, $k$, within time step $n$ as expressed below
\begin{equation}\label{eq:R}
{^n\boldsymbol{R}}^{k+1}({^n}\boldsymbol{I}^{k+1}_{CFD}) = {^n\boldsymbol{R}}^{k}({^n}\boldsymbol{I}^{k}_{CFD}) + {^n\boldsymbol{J}}^k\left({^n\Delta} \boldsymbol{I}^{k}_{CFD} \right) = \boldsymbol{0},
\end{equation}
where $\boldsymbol{J}$ is the Jacobian matrix which contains the partial derivatives of the residual vector $\boldsymbol{R}= [r_1, r_2, ..., r_m]$ with respect to the terms of the CFD input vector $\boldsymbol{I}_{CFD} = [I_1, I_2, ..., I_m]$. The Jacobian at the $n^{th}$ time step and $k^{th}$ iteration is given by
\begin{equation}\label{eq:J}
^n\mathbf{J}^k =
\begin{bmatrix}
\frac{{^n}\partial r_1^{k}}{{^n}\partial I_1^{k}} & 
\frac{{^n}\partial r_1^{k}}{{^n}\partial I_2^{k}} & 
... &
\frac{{^n}\partial r_1^{k}}{{^n}\partial I_m^{k}} \\[1ex]

\frac{{^n}\partial r_2^{k}}{{^n}\partial I_1^{k}} & 
\frac{{^n}\partial r_2^{k}}{{^n}\partial I_2^{k}} & 
... &
\frac{{^n}\partial r_2^{k}}{{^n}\partial I_m^{k}} \\[1ex]

\vdots&
\vdots& 
\ddots &
\vdots \\[1ex]

\frac{{^n}\partial r_m^{k}}{{^n}\partial I_1^{k}} & 
\frac{{^n}\partial r_m^{k}}{{^n}\partial I_2^{k}} & 
... &
\frac{{^n}\partial r_m^{k}}{{^n\partial I_m^{k}}}
\end{bmatrix}.
\end{equation}

The unknown terms of the Jacobian are approximated by finite differences:
\begin{equation}\label{eq:FD}
\frac{{^n}\partial r_i^k}{{^n}\partial I_j^k} \approx \frac{{^n}r_i^k - {^n}r_i^{k-1}}{{^n}I_j^k - {^n}I_j^{k-1}}.
\end{equation}

Once the Jacobian is known, $^n\Delta \boldsymbol{I}^{k}_{CFD}$ is calculated based on Equation~\ref{eq:R} and is added to the CFD input vector of the current iteration $k$ to get the input vector for the next iteration $k+1$:
\begin{equation}\label{eq:update}
{^n}\boldsymbol{I}^{k+1}_{CFD} = {^n}\boldsymbol{I}^{k}_{CFD} + {^n}\Delta \boldsymbol{I}^{k}_{CFD}.
\end{equation}

This method is also known as the interface Quasi-Newton method. The computed Jacobian matrix for a certain time step $n$ obtained for the first iteration $k=0$ can be used for several following coupling iterations as long as the following condition is met
\begin{equation}\label{eq:Rcriteria}
\|{^n}\boldsymbol{R}^k\| < \frac{\|{^n}\boldsymbol{R}^0\|}{10},
\end{equation}
\noindent where 10 is a heuristic value that is introduced to assure a quick convergence~\cite{toti2018coupled}. If the condition is not met, the Jacobian needs to be recomputed. For a more detailed description of the procedure to create and to evaluate the Jacobian the reader is referred to~\cite{toti2018coupled}.

\begin{figure}[h!]
\centering
\captionsetup{justification=centering}
\includegraphics[width=0.95\linewidth]{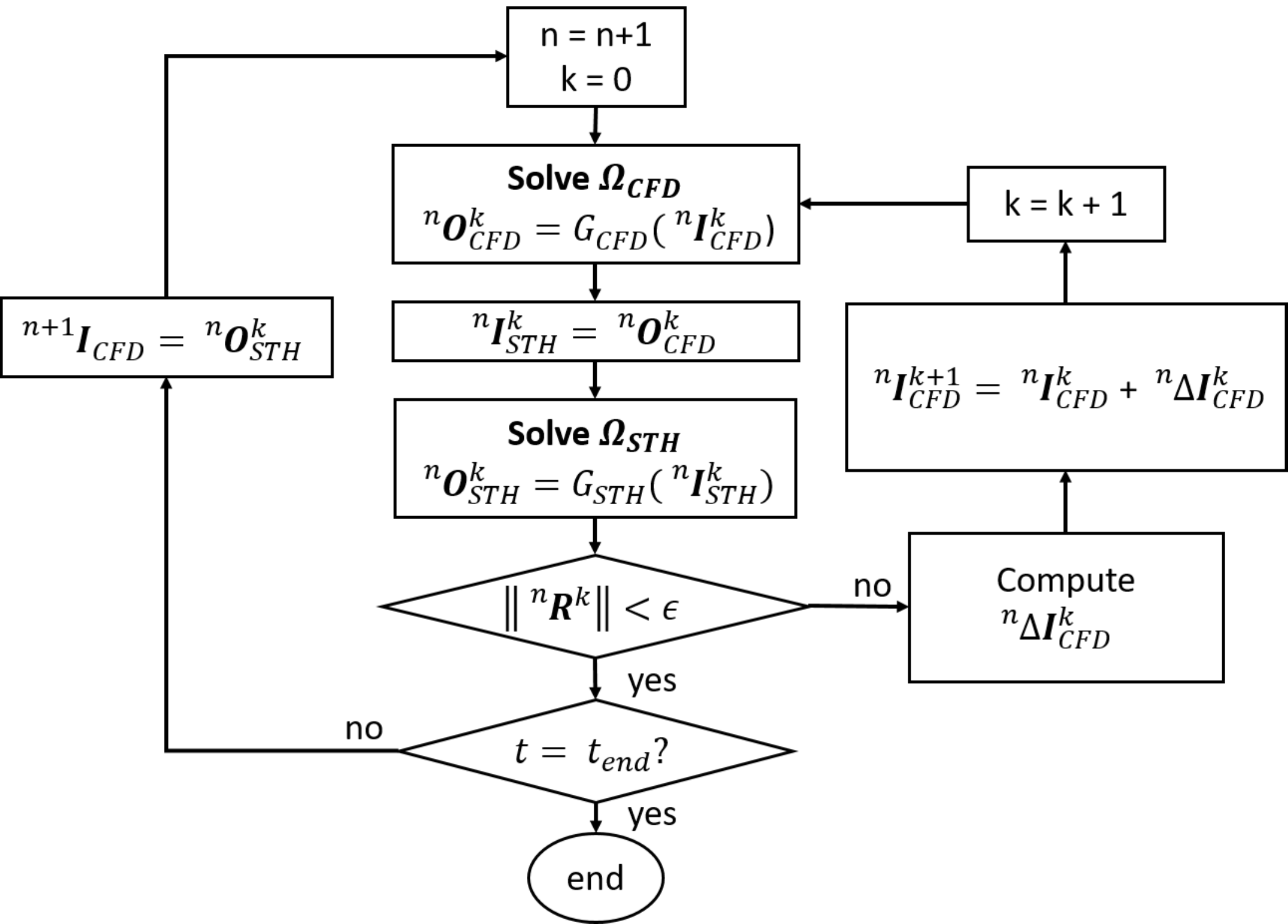}
\caption{Flowchart of the generic implicit coupling numerical algorithm with the interface Quasi-Newton algorithm. $\epsilon$ $>$ 0 is the given tolerance, $t$ denotes time and $t_{end}$ is the final simulation time.}
\label{fig:imp_scheme}
\end{figure}

Figure~\ref{fig:imp_scheme} shows a simplified flowchart of the implicit coupling algorithm.

\newpage
\section{Transport of hydraulic quantities over the coupling interfaces of a coupled RELAP5/CFD model}\label{sec:interfaces}
The transport of hydraulic quantities over the coupling interfaces of a coupled RELAP5 with OpenFOAM (RELAP5/CFD) model is explained in this section. The procedure is the same when RELAP5 is coupled with the reduced order model.

\begin{figure}[h!]
\centering
\captionsetup{justification=centering}
\includegraphics[width=1\linewidth]{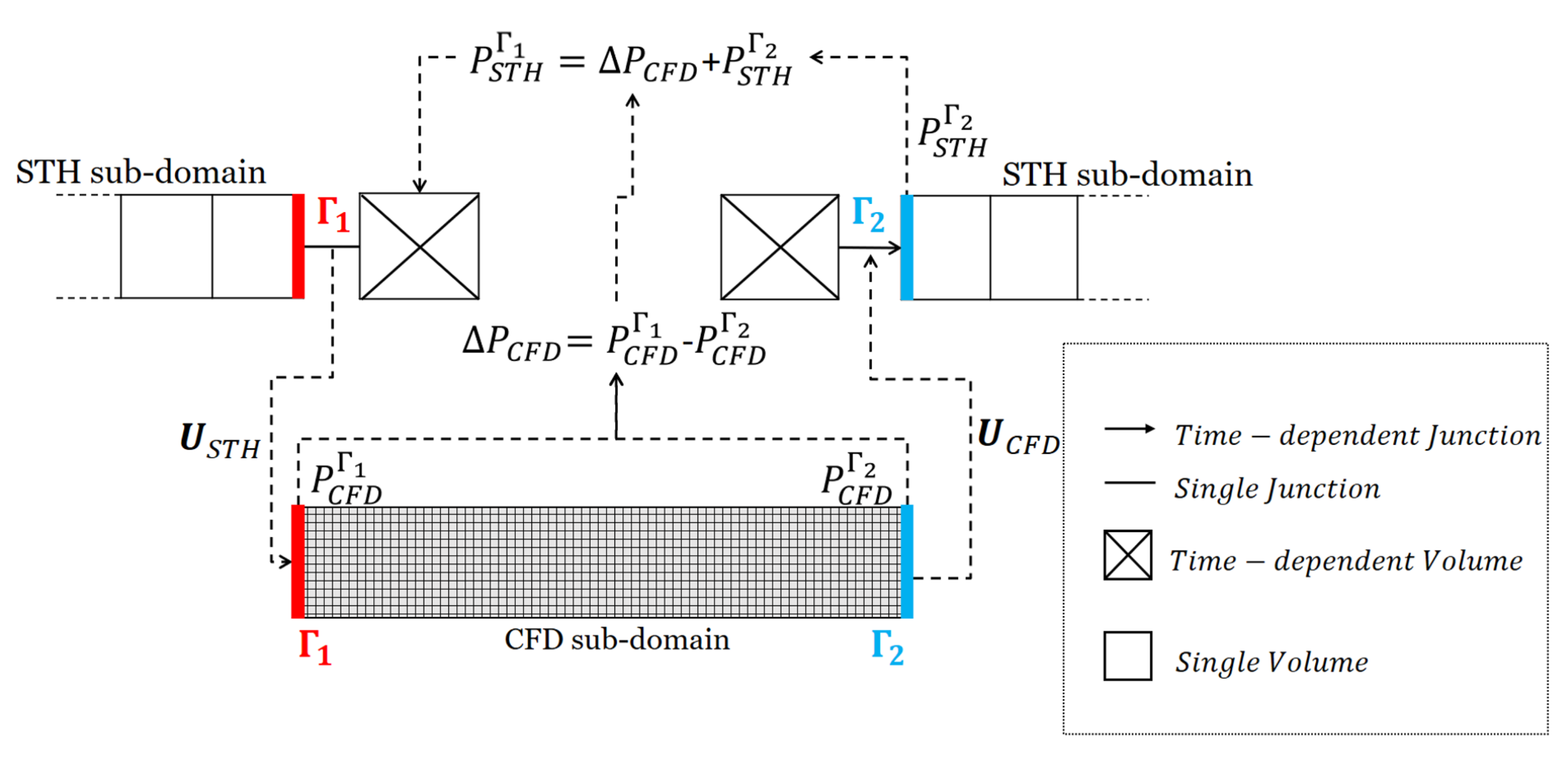}
\caption{Variable exchanges over the coupling interfaces between the STH sub-domains and the CFD sub-domain.}
\label{fig:coupled_model}
\end{figure}

As introduced previously, the coupling method is based on a domain decomposition technique. In this work, the computational
domain $\Omega$ is divided into several non-overlapping sub-domains: the STH sub-domain(s), $\Omega_{STH}$, attributed to RELAP5 and the CFD sub-domain(s), $\Omega_{CFD}$, attributed to OpenFOAM. This work is limited to simple configurations with only two interfaces, $\Gamma_1$ and $\Gamma_2$, between the STH and CFD sub-domain as depicted in Figure~\ref{fig:coupled_model}. However, the methodology is straightforward to expand to more interfaces.

Two hydraulic quantities are transported over the coupling interfaces: velocity and kinetic pressure. This is done in such a way that mass and momentum are conserved. The average velocity determined at the single junction $\boldsymbol{U}_{STH}$ of the STH sub-domain at coupling interface 1 is implemented as an uniform inlet velocity profile onto the inlet boundary of the CFD domain. At coupling interface 2, the area-averaged velocity $\boldsymbol{U}_{CFD}$ at the outlet boundary of the CFD domain is transported to the single junction of the STH sub-domain. 

\begin{figure}[h!]
\centering
\captionsetup{justification=centering}
\includegraphics[width=8cm]{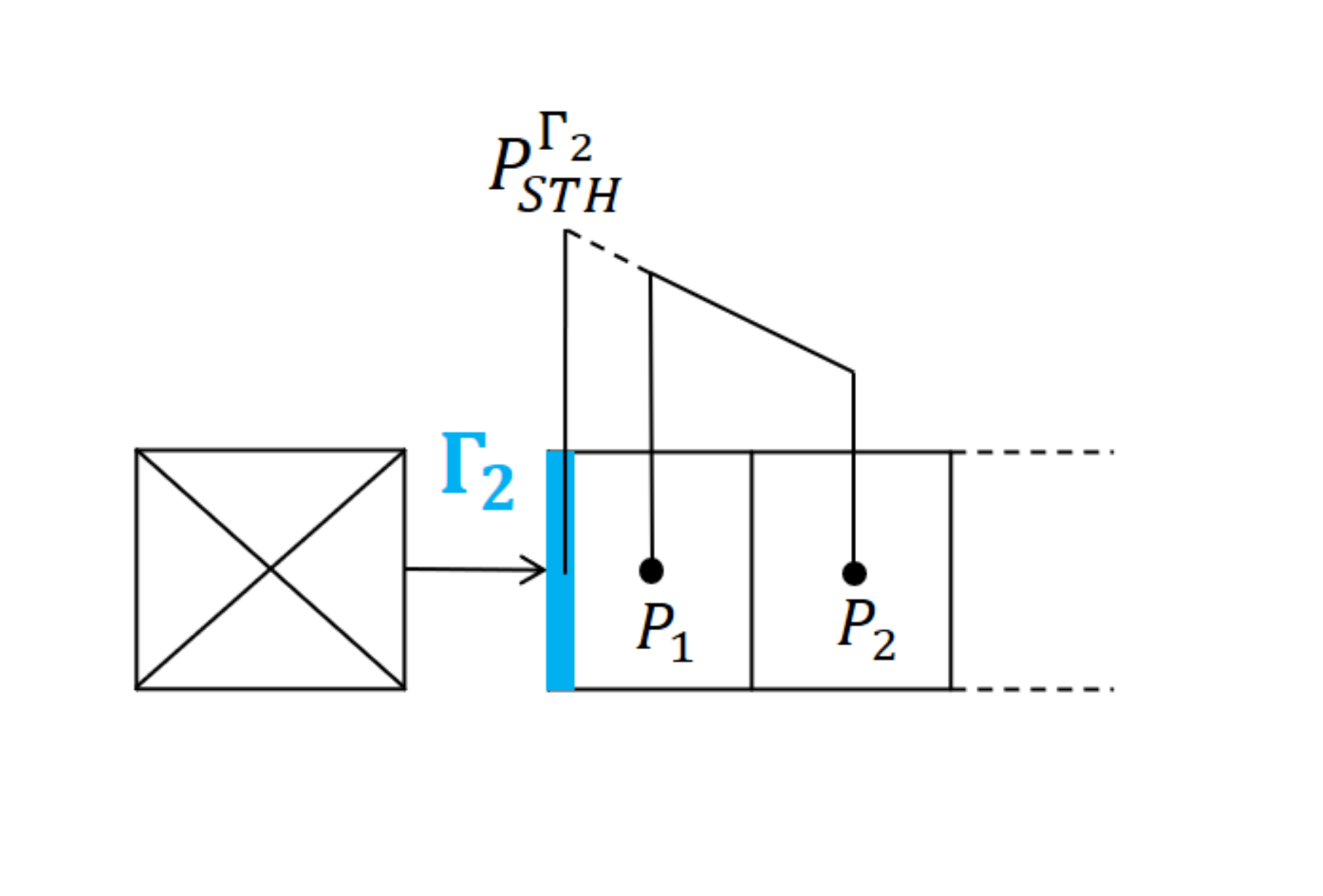}
\caption{Pressure extrapolation at coupling interface 2. The legend of Figure~\ref{fig:coupled_model} applies.}
\label{fig:p_extra}
\end{figure}

The transport of pressure over the interfaces is done differently. OpenFOAM uses the kinematic pressure, which is pressure divided by the fluid density $\rho$. Moreover, the pressure is only calculated relative to a reference level and is therefore set to 0 Pa at the outlet of the CFD domain. RELAP5 on the other hand calculates the absolute pressure at the center of all cells in the STH domain. To determine the pressure at the coupling interface 2, the volume-centered pressures from the first two neighboring cell centers of the STH domain, $P_1$ and $P_2$, as depicted in Figure~\ref{fig:p_extra}, are extrapolated to the center of the boundary of the STH domain as follows

\begin{equation}\label{eq:p_extra}
P^{\Gamma_2}_{STH}  = P_1 + \frac{P_1 - P_2}{2},
\end{equation}
\noindent where it is assumed that these neighboring cells have the same size. 

The outlet boundary of the STH sub-domain at coupling interface 1 is then updated based on the pressure determined at coupling interface 2 and the area-averaged pressure drop over the CFD sub-domain in the following way

\begin{equation}\label{eq:p_gamma_1}
P^{\Gamma_1}_{STH}  = \Delta P_{CFD} + P^{\Gamma_2}_{STH}.
\end{equation}

As the area-averaged velocity is transferred in one direction and the area-averaged pressure in the opposite direction at the coupling interfaces, transient simulations of reverse flows can also be performed using this approach~\cite{toti2018coupled}.

\section{The coupled codes' governing equations and models}\label{sec:codes}
This section presents a brief description of the best-estimate system thermal-hydraulic code RELAP5-MOD3.3. Furthermore, the governing equations that are discretized and solved with the CFD code OpenFOAM are described as those equations are projected onto a reduced basis in order to construct the ROM.

\subsection{RELAP5-MOD3.3: Thermal-hydraulic modeling}\label{sec:RELAP5}
The RELAP5-MOD3.3 code~\cite{relap5mod} is developed at the Idaho National Engineering \& Environmental Laboratory (INEEL) for the U.S. Nuclear Regulatory Commission. The code makes use of a two-fluid model in 1D form. The computational domain is subdivided in volumes that are joined by junctions. In the RELAP5 approach, the equations of mass and energy are solved in the control volumes and momentum equations are solved in the junction components i.e.\ across the two volumes. Therefore, quantities that come from the solution of the mass and energy equations, like pressure and temperature, are evaluated at the center of the nodes. Instead, quantities like velocity and mass flow rate, that come from the solution of the momentum equation are evaluated at the interface between two adjoining volumes (junction).


\subsection{OpenFOAM 6: Reynolds-averaged Navier-Stokes equations for incompressible turbulent flow}\label{sec:OpenFOAM}
Industrial turbulent flows are often described with the Reynolds-averaged Navier-Stokes (RANS) equations. The governing unsteady RANS equations for an incompressible Newtonian flow without gravity and body forces are
\begin{align} \label{eq:FOM_mat}
\begin{cases}
\nabla \cdot \boldsymbol{U} = 0 &\mbox{in  } \Omega_{CFD} , \\
\frac{\partial \boldsymbol{U}}{\partial t} + \nabla \cdot \left( \boldsymbol{U} \otimes \boldsymbol{U}\right) =  - \nabla P + \nabla \cdot \left[(\nu + \nu_t) \left(\nabla \boldsymbol{U}+\left(\nabla \boldsymbol{U}^T\right)\right)\right]  &\mbox{in  } \Omega_{CFD},  
\end{cases}
\end{align}
\noindent where $\boldsymbol{U}$ are the time-averaged values for velocity and $P$ is the time-averaged kinematic pressure, which is pressure divided by the fluid density $\rho$, $t$ denotes time and $\nu$ is the kinematic viscosity. 
Often one or two equation turbulence models are used to model incompressible turbulent flows. This work uses the $k$-$\epsilon$ turbulence model~\cite{launder1974application}, according to the previous work done by Toti et al.~\cite{toti2018coupled}. In this model the eddy viscosity, $\nu_t$, is a function of the two variables $k$ and $\epsilon$, which stand, respectively, for the turbulence kinetic energy and the turbulence dissipation rate.  


OpenFOAM uses a finite volume discretization (FV) method for which the computational domain is broken into smaller regions that are called control volumes~\cite{moukalled2016finite}. In this work, the RANS equations~\ref{eq:FOM_mat} are discretized and solved using a PIMPLE~\cite{ferziger2002computational} algorithm for the pressure-velocity coupling, which is a combination of SIMPLE~\cite{patankar1983calculation} and PISO~\cite{issa1986solution}.

\section{POD-Galerkin reduced order model for incompressible turbulent flow}\label{sec:ROM}
The reduced order model for the full order CFD code is constructed using a POD-Galerkin technique. POD stands for Proper Orthogonal Decomposition and is used to reduce the dimensionality of a system by transforming the original set of $N_x$ degrees of freedom into a new set of $N_r$ degrees of freedom, so-called modes, where $N_r$ $<$ $N_x$. These modes are ordered in such a way that the first few modes retain most of the energy present in the original solution~\cite{Lassila}. For more details about POD and other reduced basis techniques, the reader is referred to~\cite{hesthaven2016certified,quarteroni2015reduced,rozza2007reduced,chinesta2011short}. 

Flow field solutions obtained by solving the unsteady RANS equations, so-called snapshots, are collected at certain time instances. As the finite volume discretization is used on collocated grids, the variables velocity and pressure are known at discrete points in the spatial domain, which is at center of the control volumes~\cite{ferziger2002computational}. POD assumes that these solutions can be expressed as a linear combination of spatial modes multiplied by time-dependent coefficients. The $L^2$-norm is preferred for discrete numerical schemes~\cite{Stabile2017CAF,busto2020pod} with ${\left( \cdot,\cdot\right)_{L^2(\Omega_{CFD})}}$ the $L^2$-inner product of the fields over the domain $\Omega_{CFD}$. As the POD modes are orthonormal to each other, ${\left( \boldsymbol{\varphi}_i,\boldsymbol{\varphi}_j\right)_{L^2(\Omega_{CFD})}} = \delta_{ij}$ holds, where $\delta$ is the Kronecker delta.

For the velocity and pressure fields, the approximations are given, respectively, by
\begin{equation}\label{eq:approxU}
\boldsymbol{U}(\boldsymbol{x},t) \approx \boldsymbol{U}_r = \sum\limits_{i=1}^{N_r^U} \boldsymbol{\varphi}_i(\boldsymbol{x})a_{i}(t),
\end{equation}
\begin{equation}\label{eq:approxP}
P(\boldsymbol{x},t) \approx P_r = \sum\limits_{i=1}^{N_r^P} \chi_i(\boldsymbol{x})b_{i}(t), 
\end{equation}
\noindent where $\boldsymbol{\varphi}_i$ and $\chi_i$ are the modes of the velocity and pressure, and respectively $a_i$ and $b_i$ the corresponding time-dependent coefficients. $N_r^U$ is the number of velocity modes and $N_r^P$ is the number of pressure modes.

The above assumptions are extended to the eddy viscosity fields, $\nu_t$ as follows
\begin{equation}\label{eq:approxnut}
\nu_t(\boldsymbol{x},t) \approx \nu_{t_r} = \sum\limits_{i=1}^{N_r^{\nu_t}} \eta_i(\boldsymbol{x})c_{i}(t),
\end{equation}
with $\eta_i(\boldsymbol{x})$ the eddy viscosity modes and $c_i(t)$ the corresponding time-dependent coefficients. $N_r^{\nu_t}$ is the number of eddy viscosity modes.

The optimal POD basis space for velocity, $E^{POD}_{N_r^U}$ = span($\boldsymbol{\varphi}_1$,$\boldsymbol{\varphi}_2$, ..., $\boldsymbol{\varphi}_{N_r^U}$) is then constructed by minimizing the difference between the snapshots and their orthogonal projection onto the basis for the $L^2$-norm~\cite{quarteroni2014reduced}. This gives the following minimization problem
\begin{equation} \label{eq:min}
E^{POD}_{N_r^U} = \textrm{arg}\underset{\boldsymbol{\varphi}_1, ... ,\boldsymbol{\varphi}_{N_r^U}}{\textrm{min}} \frac{1}{N_s}\sum\limits_{n=1}^{N_s} \left\Vert \boldsymbol{U}_n(\boldsymbol{x}) - \sum\limits_{i=1}^{N_r^U} \left( \boldsymbol{U}_n(\boldsymbol{x}), \boldsymbol{\varphi_i} (\boldsymbol{x}) \right)_{L^2(\Omega^{CFD})} \boldsymbol{\varphi}_i(\boldsymbol{x})\right\Vert_{L^2(\Omega_{CFD})}^2,
\end{equation}
\noindent where $N_s$ is the number of collected velocity snapshots and $N_s$ $>$ $N_r^U$. The POD modes are then obtained from this minimization problem by solving the following eigenvalue problem on the snapshots~\cite{Stabile2017CAF,sirovich1987turbulence,Stabile2017CAIM}:
\begin{equation} \label{eq:ev} 
\boldsymbol{C^{corr}}\boldsymbol{Q}=\boldsymbol{Q}\boldsymbol{\lambda},
\end{equation}
\noindent where $C^{corr}_{ij}$ = ${\left( \boldsymbol{U}_i,\boldsymbol{U}_j\right)_{L^2(\Omega_{CFD})}}$ for $i$,$j$ = 1, ..., $N_s$ is the correlation matrix, $\boldsymbol{Q}$ is a square matrix of eigenvectors and $\boldsymbol{\lambda}$ is a diagonal matrix containing the eigenvalues. The POD modes, $\boldsymbol{\varphi}_i$, can then be constructed as follows
\begin{equation} \label{eq:POD}
\boldsymbol{\varphi}_i (\boldsymbol{x}) = \frac{1}{N_s\sqrt{\lambda_i}} \sum\limits_{n=1}^{N_s} \boldsymbol{U}_n(\boldsymbol{x}) Q_{i,n}\text{\hspace{0.5cm} for \hspace{0.1cm}}i = 1,...,N_r^U,
\end{equation}
of which the most energetic (dominant) modes are selected. The above assumptions can be extended to obtain the pressure and eddy viscosity modes.

To obtain a reduced order model, the POD is combined with the Galerkin projection, for which the momentum equations of the set of governing equations~\ref{eq:FOM_mat} are projected onto the reduced POD basis space. The following reduced system of momentum equations is then obtained 
\begin{equation}\label{eq:ROM}
\boldsymbol{M_r} \dot{\boldsymbol{a}} + \boldsymbol{a}^T\boldsymbol{C_r} \boldsymbol{a} + \boldsymbol{A_r} \boldsymbol{b} - \nu (\boldsymbol{B_r} + \boldsymbol{BT_r}) \boldsymbol{a} -  \boldsymbol{c}^T (\boldsymbol{D_r} + \boldsymbol{DT_r}) \boldsymbol{a} = \boldsymbol{0},
\end{equation}
\noindent where the 'over-dot' indicates the time derivative and
\begin{align}\label{eq:ROM_matricesM}
M_{r_{ij}} &= {\left( \boldsymbol{\varphi}_i,  \boldsymbol{\varphi}_j  \right)_{L^{2}(\Omega_{CFD})}}, \\
A_{r_{ij}}  &= {\left( \boldsymbol{\varphi}_i, \nabla \chi_j \right)_{L^{2}(\Omega_{CFD})}}, \\
B_{r_{ij}}  &= {\left( \boldsymbol{\varphi}_i, \nabla \cdot \nabla  \boldsymbol{\varphi}_j  \right)_{L^{2}(\Omega_{CFD})}}, \\
BT_{r_{ij}} &= {\left(\boldsymbol{\varphi}_i, \nabla \cdot \left(\nabla  \boldsymbol{\varphi}_j^T \right) \right)_{L^{2}(\Omega_{CFD})}}, \\
C_{r_{ijk}}  &= {\left( \boldsymbol{\varphi}_i, \nabla \cdot (\boldsymbol{\varphi}_j \otimes \boldsymbol{\varphi}_k) \right)_{L^{2}(\Omega_{CFD})}}, \\
D_{r_{ijk}}  &= {\left( \boldsymbol{\varphi}_i, \nabla \cdot \boldsymbol{\eta}_j  \nabla \boldsymbol{\varphi}_k \right)_{L^{2}(\Omega_{CFD})}}, \\
DT_{r_{ijk}}  &= {\left( \boldsymbol{\varphi}_i, \nabla \cdot \boldsymbol{\eta}_j \left( \nabla \boldsymbol{\varphi}_k^T \right) \right)_{L^{2}(\Omega_{CFD})}}. 
\end{align}

These reduced matrices and third order tensors are stored while constructing the reduced order model during a, so-called, off-line stage. More details on the POD and Galerkin projection method can be found in~\cite{Stabile2017CAF,Stabile2017CAIM,georgaka2018parametric}. 

Standard Galerkin projection-based reduced order models are unreliable when applied to the non-linear unsteady Navier-Stokes equations~\cite{Lassila,rozza2007stability,caiazzo2014numerical,Akhtar,Sirisup,bergmann2009enablers}. Often a pressure stabilization on the ROM level is required to obtain stable ROM solutions. One technique is to solve a Pressure Poisson Equation (PPE) rather than the equation for mass conservation (Equation~\ref{eq:FOM_mat}a) at the ROM level. For more details on the derivation of the PPE the reader is referred to J.-G Liu et al.~\cite{liu2010stable}. Alternative methods are possible, which are discussed in~\cite{Stabile2017CAF}.

Moreover, reduced order models for scale resolving turbulent flow simulations are often affected by the energy blow up due to the truncation of POD modes. The occurrence of unstable time behavior in the reduced order model can then be explained by the concept of the energy cascade~\cite{couplet2003intermodal}. A closure modeling between the POD-based ROM and the full order scale resolving turbulence representation is then required to improve the accuracy and instability of POD-Galerkin ROMs~\cite{wang2012proper,xie2018data,osth2014need}. ROMs based on RANS simulation typically include a closure model based on eddy viscosity at ROM level as eddy viscosity models are already present in the full order model~\cite{Versteeg,stabile2019reduced}.

Also the reduced system of equations constructed in this work, consisting of the momentum equations (Equation~\ref{eq:ROM}) together with the PPE, is not sufficient to determine all unknown coefficients (of velocity, pressure and eddy diffusivity). Therefore, the coefficients $c_i(t)$, that are used in the approximation of the eddy viscosity fields, are computed with a data-driven non-intrusive interpolation procedure using Radial Basis Functions (RBF) as described in~\cite{lazzaro2002radial}.
The advantage of approximating the eddy viscosity coefficients with RBFs is that the turbulence transport equations for $k$ and $\epsilon$ of the system of Equations~\ref{eq:FOM_mat} do not need to be projected onto the POD basis space spanned by the eddy viscosity modes. Therefore, the reduced order model is independent of the turbulence model used in the RANS simulations~\cite{hijazi2018effort}. 

The initial conditions (IC) for the reduced system of Ordinary Differential Equations (Equation~\ref{eq:ROM}) are obtained by performing a Galerkin projection of the initial conditions for the RANS simulations onto the POD basis spaces as follows 
\begin{align}\label{eq:ROM_IC1}
a_i(0) &= \left(\boldsymbol{\varphi}_i(\boldsymbol{x}),\boldsymbol{U}(\boldsymbol{x},0)\right)_{L^{2}(\Omega_{CFD})},\\
b_i(0) &= \left(\chi_i(\boldsymbol{x}),P( \boldsymbol{x},0)\right)_{L^{2}(\Omega_{CFD})},
\end{align}
\noindent for velocity and kinematic pressure, respectively.

\section{Challenges for coupling system thermal-hydraulic codes with a reduced order model}\label{sec:challenges}
In the previous section and in Section~\ref{sec:coupling}, we noted that POD-Galerkin reduced order models are, in general, sensitive to numerical instabilities~\cite{Akhtar,Sirisup,bergmann2009enablers}. This is one of the main challenges of reduced order modeling for fluid flow problems~\cite{Lassila}. Therefore, only an implicit coupling scheme is considered in this work, which is numerically more stable than the explicit coupling schemes~\cite{toti2016development}.

The Quasi-Newton coupling algorithm of Figure~\ref{fig:imp_scheme} for coupling a system thermal-hydraulic code with a CFD solver is the same for the coupling with a reduced order model. Nevertheless, there are a couple of challenges that need to be taken into account when replacing the CFD part with a ROM.  

The velocity and pressure snapshots, required for the creation of the reduced basis spaces of the ROM, are typically collected by performing high fidelity simulation. For the RELAP5/ROM coupled model, the snapshots need to be collected by performing coupled RELAP5/CFD simulations. If there are perturbations present in the coupled RELAP5/CFD solutions, it can have a detrimental effect on the performance of the RELAP5/ROM coupled model.

Moreover, reduced order models are only capable of predicting solutions for new parameter values and for long time integration if the flow features of these new cases are contained in the reduced basis spaces spanned by the POD modes~\cite{Lassila}. It is challenging to construct the optimal reduced basis spaces, especially for parametric ROMs. We will not focus on this challenge in this work. For more details on constructing the optimal reduced basis spaces we refer the reader to~\cite{Lassila,bergmann2009enablers}. 

A challenge related to the exchange of hydraulic quantities over the coupling interfaces is that non-homogeneous time-dependent boundary conditions need to be imposed in the ROM using a boundary control method~\cite{graham1999optimal1}. Another challenge is related to the accuracy of the ROM. We discuss both challenges in the subsequent subsections.

\subsection{Imposing the non-homogeneous time-dependent boundary conditions in the ROM} \label{sec:BCs}
As described in Section~\ref{sec:interfaces}, the average velocity determined at the single junction $\boldsymbol{U}_{STH}$ of the STH sub-domain at coupling interface 1 is implemented as an uniform inlet velocity profile onto the inlet boundary of the CFD domain.
Boundary conditions of the CFD domain at the coupling interfaces of a coupled RELAP5/CFD system are controlled every time step as described in Section~\ref{sec:coupling}. However, the boundary conditions are not explicitly present in the reduced momentum equations (Equation~\ref{eq:ROM}). Therefore, they cannot be controlled directly~\cite{lorenzi2016pod}. The selected approach in this work for handling the non-homogeneous Dirichlet BCs is the penalty method~\cite{graham1999optimal1}. The aim of the penalty method is to enforce the BCs in the ROM with a penalty factor $\tau$~\cite{Sirisup}. The Dirichlet BC, $\boldsymbol{U}_{BC}$, for velocity is implemented in the momentum equations as follows
\begin{equation} \label{eq:mom_penalty}
\frac{\partial \boldsymbol{U}}{\partial t} + \nabla \cdot \left( \boldsymbol{U} \otimes \boldsymbol{U}\right) + \nabla P - \nabla \cdot \left[(\nu + \nu_t)\left(\nabla \boldsymbol{U}+\left(\nabla \boldsymbol{U}^T\right)\right) \right]  + \tau\Gamma_{BC}(\boldsymbol{U} - \boldsymbol{U_{BC}}) = 0.
\end{equation} 
where $\Gamma_{BC}$ is a null function except on the boundary where the Dirichlet boundary condition is imposed~\cite{Sirisup,lorenzi2016pod}. Substituting the approximated expansions (Equations~\ref{eq:approxU}-\ref{eq:approxnut}) for the fields into Equation~\ref{eq:mom_penalty} and applying the Galerkin projection results in the following set of ODEs
\begin{equation}\label{eq:ROM_mom2}
\boldsymbol{M_r} \dot{\boldsymbol{a}} + \boldsymbol{a}^T\boldsymbol{C_r} \boldsymbol{a} + \boldsymbol{A_r} \boldsymbol{a} - \nu (\boldsymbol{B_r} + \boldsymbol{BT_r}) \boldsymbol{a} - \boldsymbol{c}^T (\boldsymbol{D_r} + \boldsymbol{DT
_r}) \boldsymbol{a}  + \tau\left(\boldsymbol{u}_{BC}\boldsymbol{E_r} - \boldsymbol{F_r}\boldsymbol{a}\right) =0,
\end{equation}

\noindent where 
\begin{align}\label{eq:ROM_matricesN1}
E_{r_{i}}  &= {\left( \boldsymbol{\varphi}_i, \boldsymbol{\zeta}  \right)_{L^{2}(\Gamma)}}, \\
F_{r_{ij}}  &= {\left( \boldsymbol{\varphi}_i, \boldsymbol{\varphi}_j  \right)_{L^{2}(\Gamma)}},
\end{align}
\noindent with $\boldsymbol{\zeta}$ a unit field and $\Gamma$ the boundary of the CFD domain.

The penalty factor is usually tuned with a sensitivity analysis~\cite{lorenzi2016pod,graham1999optimal1,kalashnikova2012efficient}. However, if $\tau$ tends to zero, the BCs are not enforced and if the factor tends to infinity the ROM becomes ill-conditioned~\cite{lorenzi2016pod}.

\subsection{Relative error}
Reduced order models contain a lower number of degrees of freedom than the high fidelity models due to a truncation of the POD modes. That way, they are computationally more efficient, but have generally a lower accuracy than the high fidelity models~\cite{Lassila,grepl2007efficient}. 

To determine the accuracy of the coupled model of which the CFD part is replaced by a ROM, the solutions need to be compared with those of the coupled RELAP5/CFD model.

In this work, the accuracy of the RELAP5/ROM coupled model is determined by calculating the relative $L^2$-error for each time step, $t$, between the RELAP5/CFD solutions, $X_{CFD}$, and the fields obtained by performing coupled RELAP5/ROM simulations, $X_{ROM}$. This so-called relative prediction error is defined as 
\begin{equation}\label{eq:l2_prediction}
\|e\|_{L^2(\Omega_{CFD})}(t) = \frac{\|X_{CFD}(t)-X_{ROM}(t)\|_{L^{2}(\Omega_{CFD})}}{\|X_{CFD}(t) \|_{L^{2}(\Omega_{CFD})}}.
\end{equation}
\noindent where $X$ represents the velocity or pressure fields.

We compare the prediction errors with the basis projection error, which acts as a lower error bound for the reduced order model. The basis projection error, $\|\hat{e}\|_{L^2(\Omega_{CFD})}$, is defined as the relative $L^2$-error between the RELAP5/CFD solutions, $X_{CFD}$, and the projected fields, $X_r$, which are obtained by the $L^2$-projection of the snapshots onto the POD bases: 
\begin{equation}\label{eq:l2_projection}
\|\hat{e}\|_{L^2(\Omega_{CFD})}(t) = \frac{\|X_{CFD}(t)-X_{r}(t)\|_{L^{2}(\Omega_{CFD})}}{\|X_{CFD}(t) \|_{L^{2}(\Omega_{CFD})}}.
\end{equation}
In practice, the prediction error is larger than the projection error for a single parameter point.

\section{Set-up numerical test cases} \label{sec:setup} 
In this section, the set-ups for three different configurations are described: the open pipe flow test, the open pipe flow reversal test and the closed pipe flow test. All tests are carried out for single-phase water flow with kinematic viscosity $\nu$ = 1.0$\cdot$\num{e-6} m$^2$/s. 

\begin{figure}[h!]
\centering
\captionsetup{justification=centering}
\includegraphics[width=9cm]{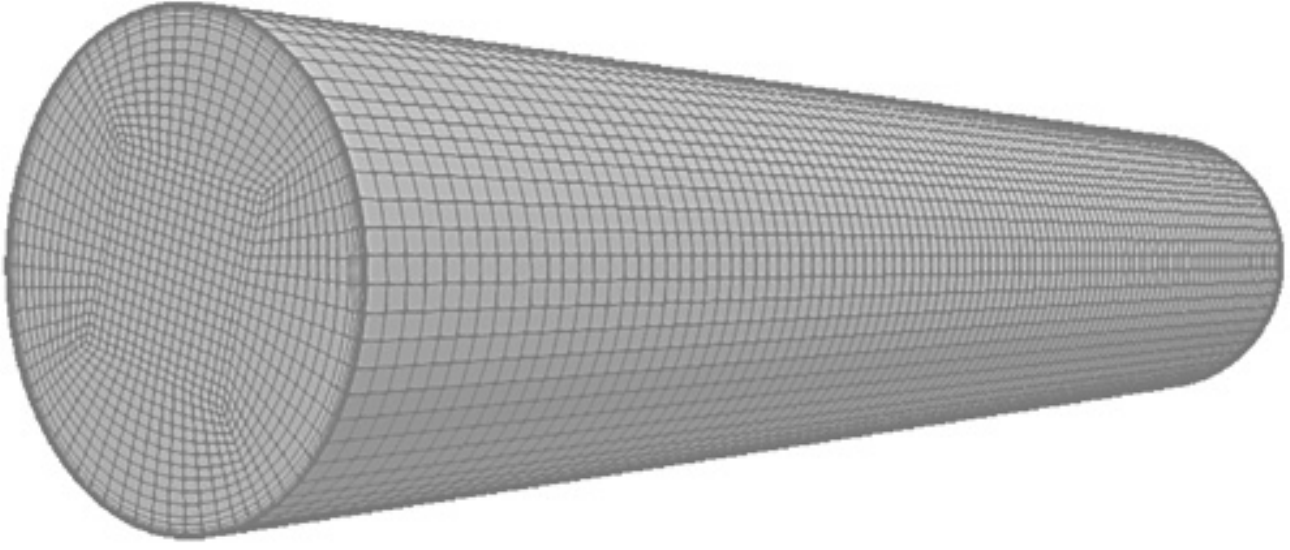}
\caption{Mesh of the CFD sub-domain consisting of a circular pipe of length $L_{CFD}$ and diameter $D$.}
\label{fig:mesh}
\end{figure}

For the coupled models, the computational domain is divided into a CFD sub-domain and an STH sub-domain. For all configurations, the CFD sub-domain consists of a circular pipe of length $L_{CFD}$ = 0.5 m and diameter $D$ = 0.1 m. A mesh with 145945 hexahedral cells is constructed onto the three-dimensional domain, as depicted in Figure~\ref{fig:mesh}. As the CFD sub-domain and mesh are kept unchanged, the coupling procedure for coupling RELAP5 with the reduced order model is the same as for the RELAP5/CFD coupled model. 

The unsteady RANS equations (Equation~\ref{eq:FOM_mat}) are discretized and solved by the finite volume method with ITHACA-FV~\cite{ITHACA}, which is an open source C++ library based on the finite volume solver OpenFOAM~\cite{Jasak}. In this work, the libraries of OpenFOAM 6 are used. The spatial discretization is performed with linear interpolation schemes and the temporal discretization is treated using a first order implicit differencing scheme. The calculation of the POD modes, the Galerkin projection of the RANS solutions on the reduced subspace and the ROM simulations are also carried out with ITHACA-FV. For more details on the code, the reader is referred to~\cite{Stabile2017CAF,Stabile2017CAIM,ITHACA}. 

All stand-alone STH models of the computational domains are constructed with RELAP5-MOD3.3. The reduced order CFD models are constructed according to the methodology described in section~\ref{sec:ROM}. The velocity and pressure snapshots needed for the creation of the reduced subspaces are collected by performing a coupled RELAP5/CFD simulation for a certain parameter set. This is required for each of the three flow configurations. Moreover, the coupled RELAP5/CFD model is first evaluated against the corresponding STH stand alone model in order to evaluate the implicit coupling methodology. Thereafter, the coupled RELAP5/ROM models are tested and compared with the coupled RELAP5/CFD models. 

All simulations are run on a single Intel\textsuperscript{\tiny\textregistered} Xeon\textsuperscript{\tiny\textregistered} GOLD 5118 @ 2.30GHz core. 

\subsection{Open pipe flow test} \label{sec:open_pipe} 
\begin{figure}[h!]
\centering
\begin{subfigure}[b]{0.9\textwidth}
	\includegraphics[width=1\linewidth]{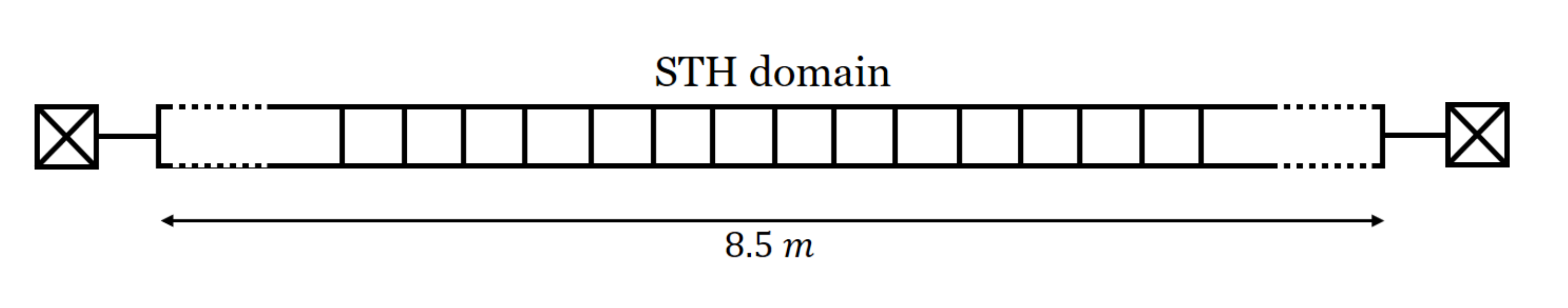}
\end{subfigure}
\begin{subfigure}[b]{0.9\textwidth}
	\includegraphics[width=1\linewidth]{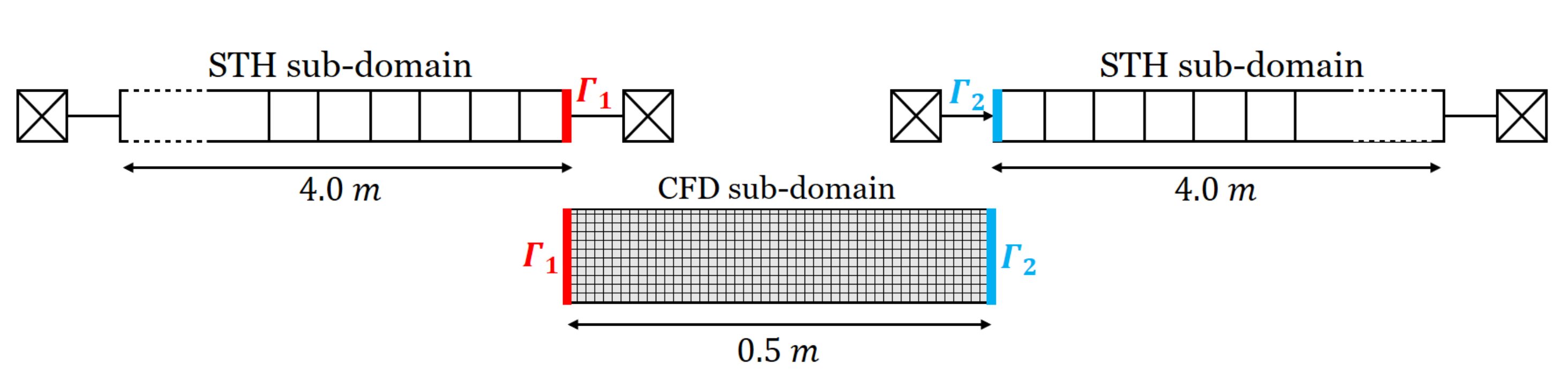}
\end{subfigure}%
\caption{RELAP5 stand-alone nodalization of the open pipe flow configuration (top) and the domain decomposition of the open pipe flow configuration for the coupled RELAP5/CFD model (bottom). The legend of Figure~\ref{fig:coupled_model} applies.}
\label{fig:sth_stand_alone_setup_sth_of}
\end{figure}

The simple open pipe configuration consists of a circular straight pipe with length $L$ = 8.5 m and internal diameter $D$ = 0.1 m. The pipe is split in three parts; the beginning and the ending parts have both have a length $L_{STH-1}$ = $L_{STH-2}$ = 4.0 m and the middle part has length $L_{CFD}$ = 0.5 m. Figure~\ref{fig:sth_stand_alone_setup_sth_of} shows a sketch of the set-up for a STH stand-alone simulation at the top. In the same figure, the set-up of the coupled RELAP5/CFD model is shown at the bottom. The beginning and ending parts of the STH domain are divided in 10 volumes of 0.4 m. The middle is either modeled with a finer mesh of 20 equally sized volumes of 0.0025 m for the STH stand-alone simulations or assigned to the CFD code for the coupled simulations.

The fluid is initially at rest and driven by an abrupt pressure difference, $\Delta P$, of 0.20 bar applied over the whole pipe at $t$ = 0 s. The total time of simulation is 10 s. The inlet and outlet boundary conditions of the STH sub-domains are set by time-dependent volumes. A previous study by Toti at al.~\cite{toti2018coupled} showed that accurate results for velocity and pressure is obtained for a time step of 0.1 s in case of implicit coupling. Therefore, the coupled simulations are performed with this time step. 

The coupling methodology is evaluated by comparing the time evolution of the mass flow rate and the area-averaged pressure at coupling interface $\Gamma_2$ obtained with RELAP5/CFD and the stand-alone RELAP5 simulations for the pressure drop of 0.20 bar. Snapshots of the velocity and pressure fields that are calculated at the CFD sub-domain are collected every 0.1 s during the coupled RELAP5/CFD simulation. Thus, 100 snapshots are collected in total that are used for the construction of the RELAP5/ROM model according to section~\ref{sec:ROM}. 

The coupled RELAP5/ROM model is then tested for the same pressure difference and four new conditions, namely $\Delta P $ = 0.10, 0.15, 0.21 and 0.23 bar. As described in Section~\ref{sec:BCs}, the uniform velocity boundary value at the inlet is enforced in the ROM with a penalty method. In this work, the penalty factor is set to 1.0. The RELAP5/ROM results are compared with corresponding coupled RELAP5/CFD results.

\subsection{Open pipe flow reversal test} \label{sec:reversal} 
Using the same set-up as for the open pipe flow test case, both the RELAP5/CFD and the RELAP5/ROM models are tested for sudden flow reversal. Initially, the absolute pressure at the inlet of the STH domain is set to 1.40 bar while the outlet pressure is set to 1.20 bar. Thus, the total pressure drop over the whole pipe is 0.20 bar. Between $t$ = 9 s and $t$ = 13 s, the pressure at the inlet is decreased linearly up to 1.0 bar and the simulation is run up to $t$ = 25 s. Once the pressure at the inlet is lower than the pressure at the outlet, the fluid eventually starts flowing in the opposite direction. This is tested for STH stand alone, RELAP5/CFD and RELAP5/ROM. As done for the open pipe flow test, snapshots are collected every 0.10 s. Moreover, the coupled RELAP5/ROM model is tested for pressure drops of 0.10, 0.15, 0.21 and 0.23 bar.

Furthermore, the ROM performance for long time integration is tested with this test case. 100 velocity, pressure and turbulence viscosity snapshots are collected during the first 10 seconds of simulation time. The obtained RELAP5/ROM model, after performing POD onto the snapshots, is used to simulate the whole transient up to $t$ = 25 s and the results are compared with an additional RELAP5/ROM for which 250 snapshot were collected during the whole simulation time.

\subsection{Closed pipe flow test} \label{sec:closed_pipe} 
\begin{figure}[h!]
\centering
\includegraphics[width=0.8\linewidth]{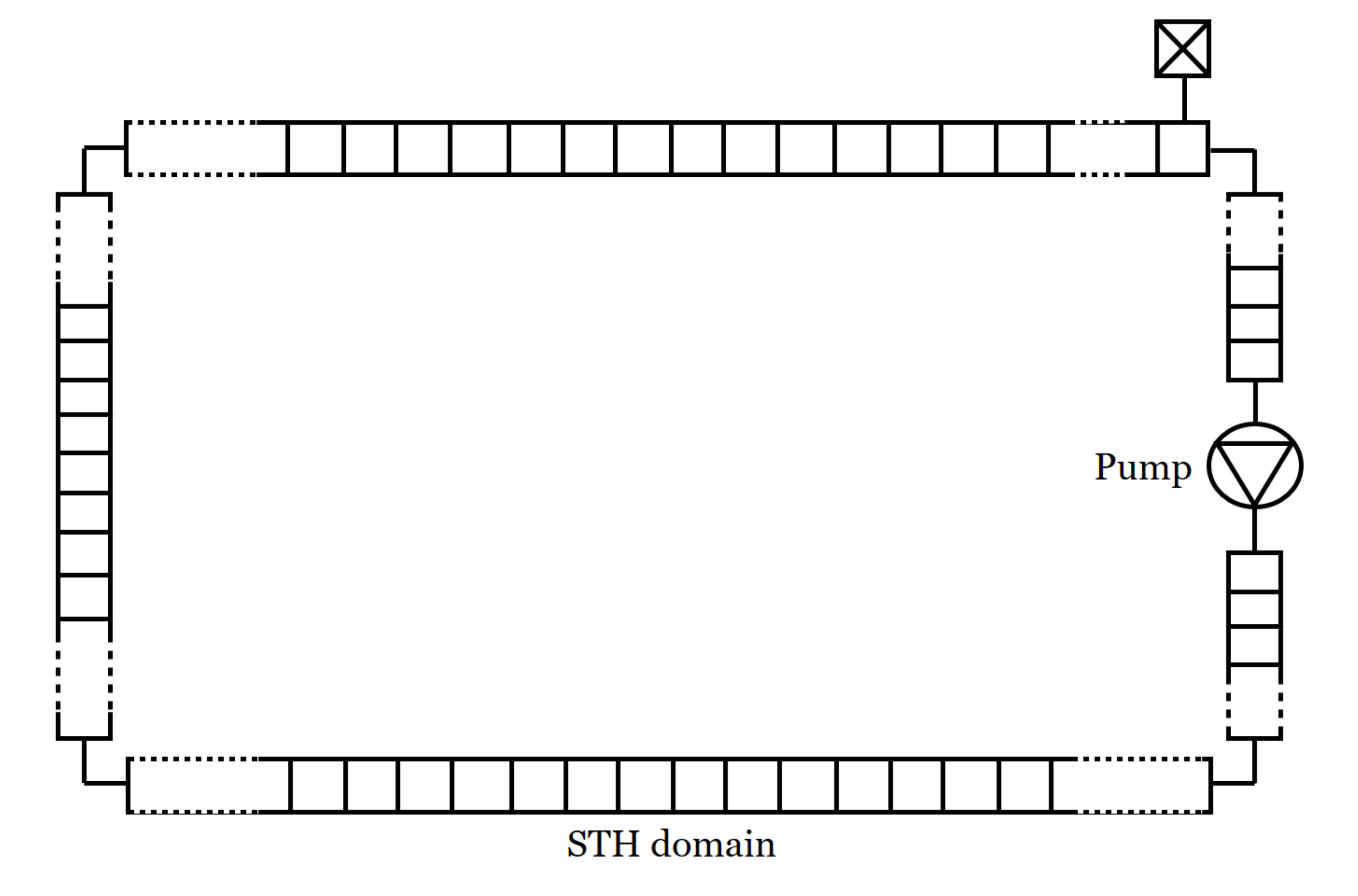}
\caption{Set-up of the STH stand-alone normalization of the closed pipe flow configuration. The legend of Figure~\ref{fig:coupled_model} applies.}
\label{fig:setup_sth_of_closed}
\end{figure}

\begin{figure}[h!]
\centering
\includegraphics[width=0.8\linewidth]{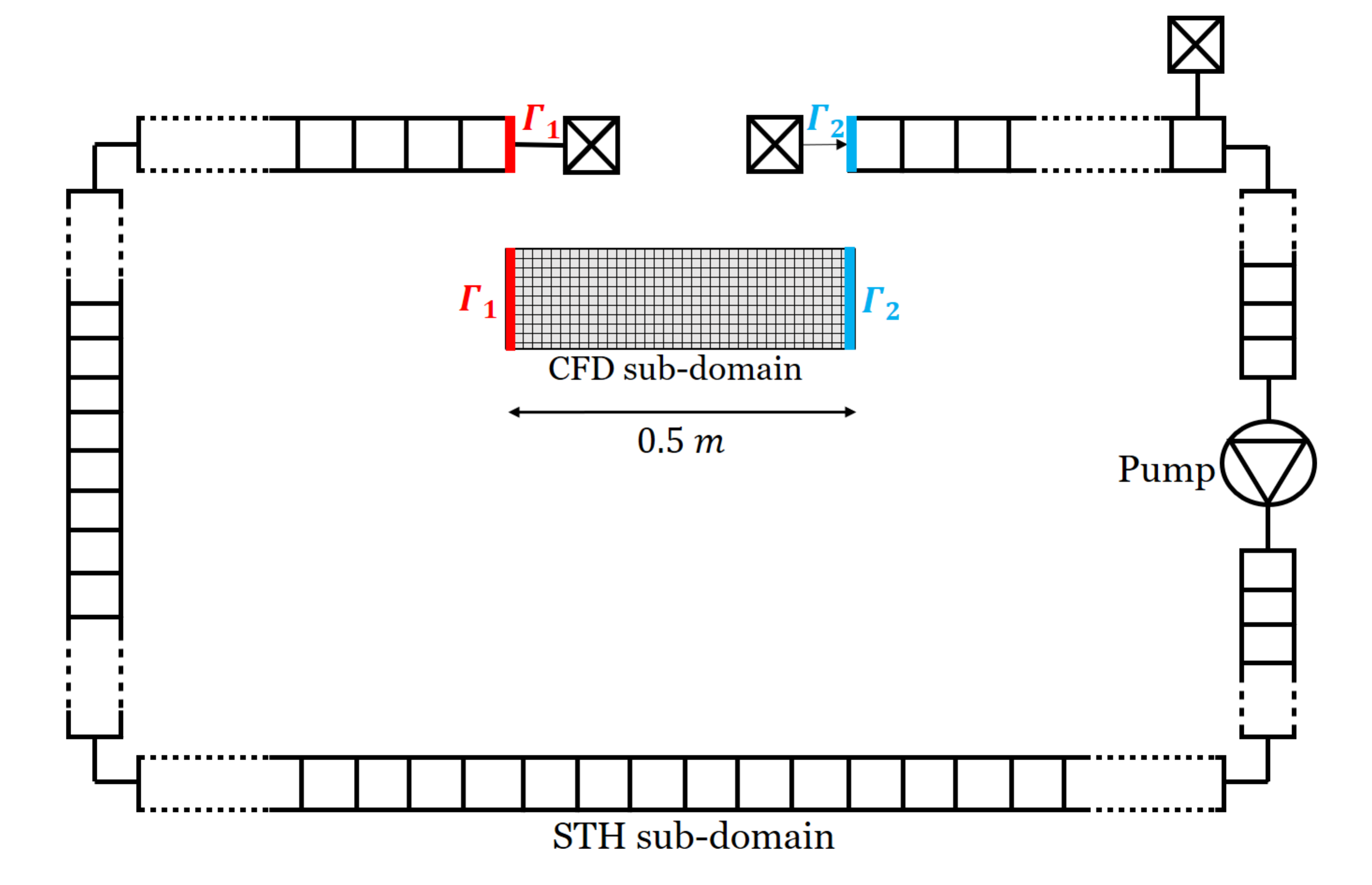}
\caption{Set-up of the closed pipe flow configuration for the domain decomposition coupled RELAP5/CFD model. The legend of Figure~\ref{fig:coupled_model} applies.}
\label{fig:setup_sth_of_closed_cfd}
\end{figure}

The last configuration consists of a closed loop. The STH configuration of the open pipe flow test is extended with a circulation pump and an expansion tank as shown in Figure~\ref{fig:setup_sth_of_closed}. Figure~\ref{fig:setup_sth_of_closed_cfd} shows the same configuration with the domain decomposition for the coupled simulations. The two vertical legs of the loop are 1.6 m long while the horizontal legs are 8.5 m long. The top horizontal leg is split similarly to the open pipe flow test case. In this configuration, the transient is initiated by the start of the pump causing again a mass flow ramp. The pump reaches its nominal speed after 5 seconds of simulation time. The total simulation time is 10 s. A coupled RELAP5/CFD simulation is performed with the nominal speed of the rotor set to 100 rad/s and snapshots are collected every 0.10 seconds. Coupled RELAP5/ROM simulations are also performed for a nominal rotor speed of 80 rad/s, 90 rad/s and 110 rad/s .



\newpage
\section{RESULTS}\label{sec:results}
For each of the flow configurations, the coupled model is first evaluated against the corresponding STH stand alone model in order to evaluate the implicit coupling methodology. Thereafter, the coupled RELAP5/ROM models are tested and compared with the coupled RELAP5/CFD models.

\subsection{Open pipe flow test} \label{sec:res_open_pipe} 
\begin{figure}[h!]
\centering
\includegraphics[width=0.7\linewidth]{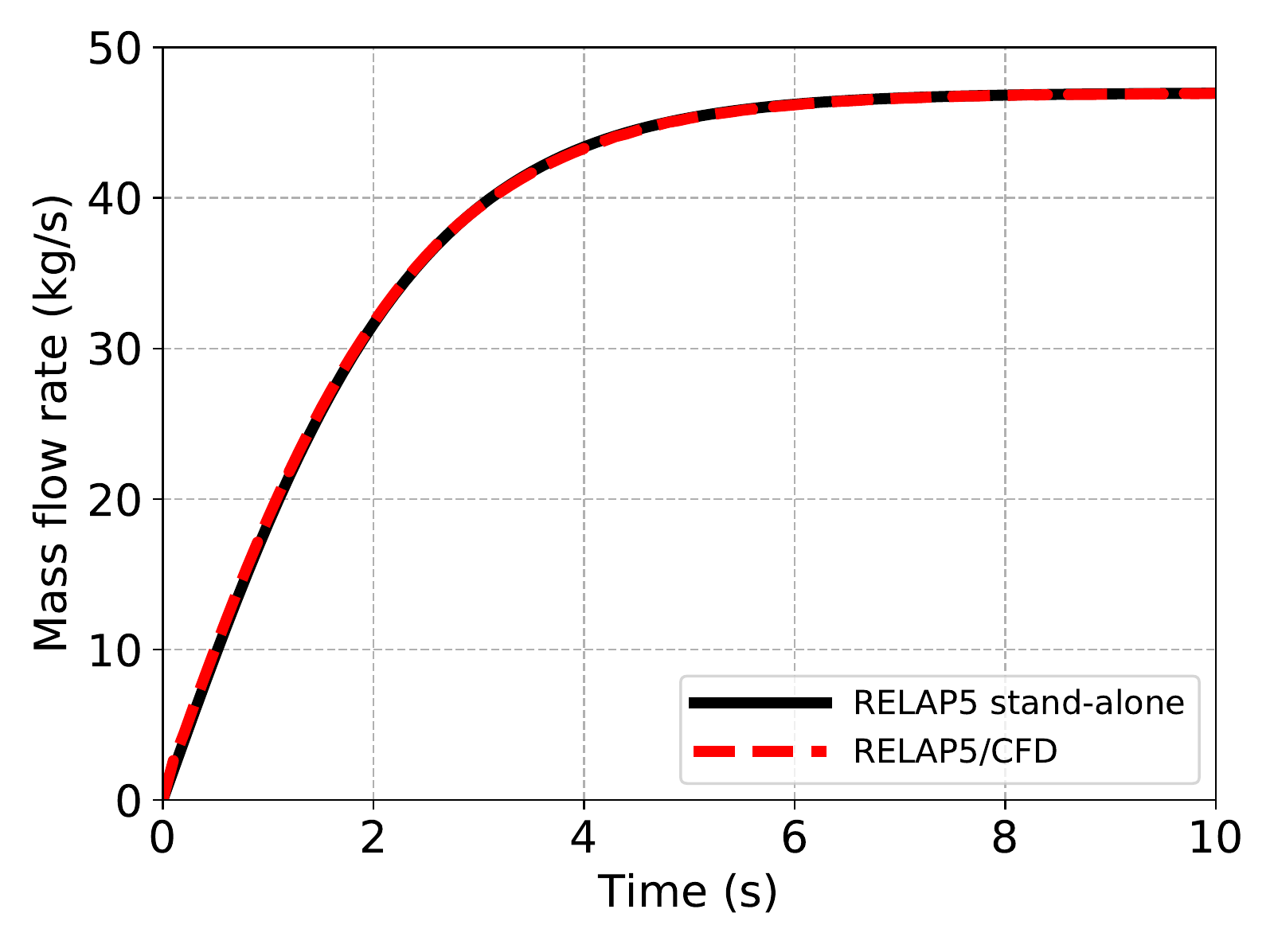}
\caption{Time evolution of the mass flow rate through the pipe at interface 2 in the abrupt pressure difference transient ($\Delta P$ = 0.20 bar) for an open pipe flow configuration. }
\label{fig:res_open_phi}
\end{figure}

\begin{figure}[h!]
\centering
\includegraphics[width=0.7\linewidth]{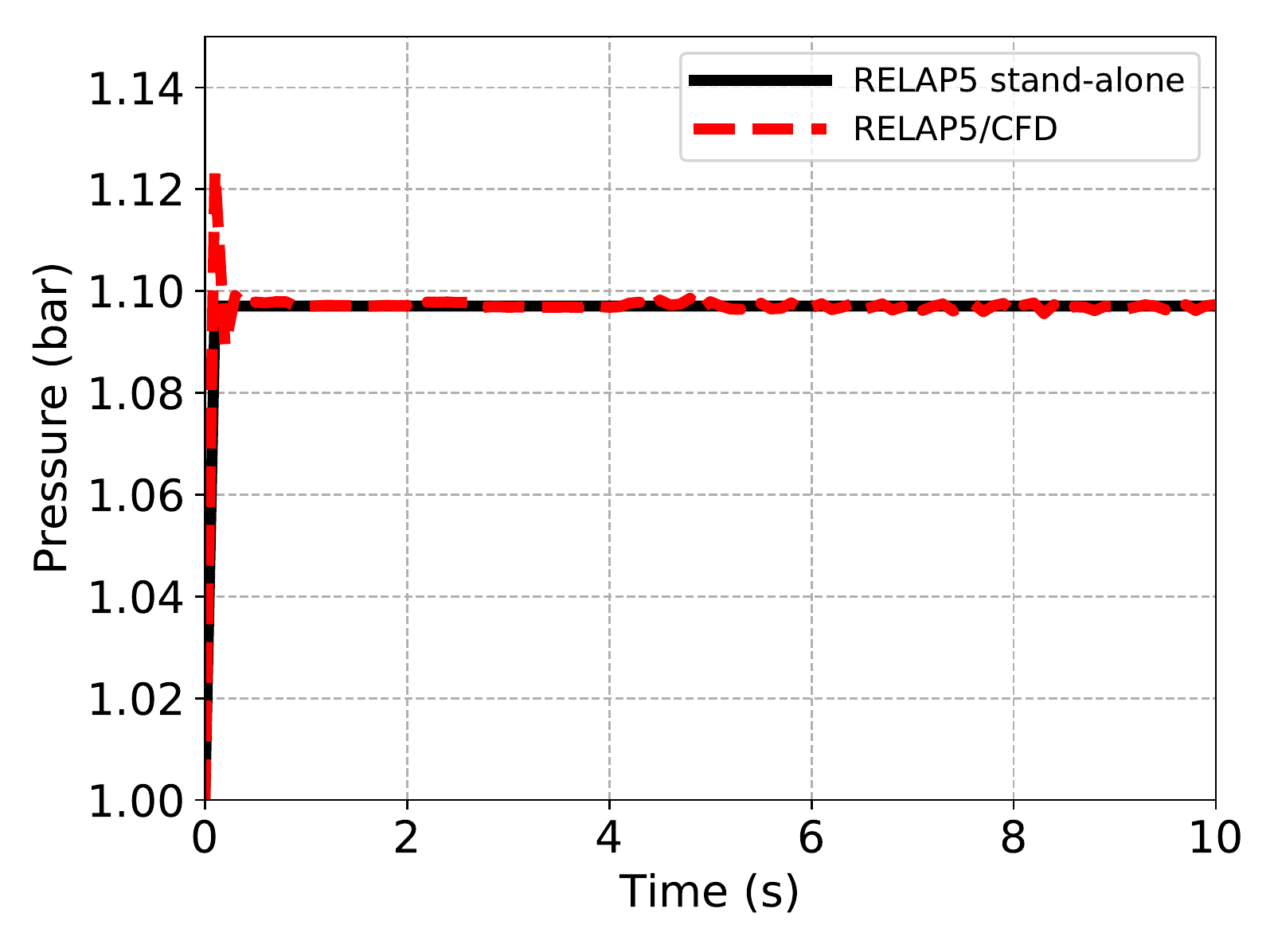}
\caption{Time evolution of the pressure at interface 2 in the abrupt pressure difference transient ($\Delta P$ = 0.20 bar) for an open pipe flow configuration. }
\label{fig:res_open_p}
\end{figure}

The coupled RELAP5/CFD system is tested for a pressure drop of 0.20 bar over the open pipe. The results are compared with the results of the RELAP5 stand alone simulation. Figures~\ref{fig:res_open_phi} and~\ref{fig:res_open_p} show the time evolution of the mass flow rate at interface $\Gamma_2$ and the area-averaged pressure at the same interface, respectively. Pressure oscillations are present at the beginning of the simulation, but they dissolve as the simulation time proceeds. The oscillations in pressure do not result in oscillations of the mass flow rate. 

\begin{figure}[h!]
\centering
\captionsetup{justification=centering}
\includegraphics[width=1.\textwidth]{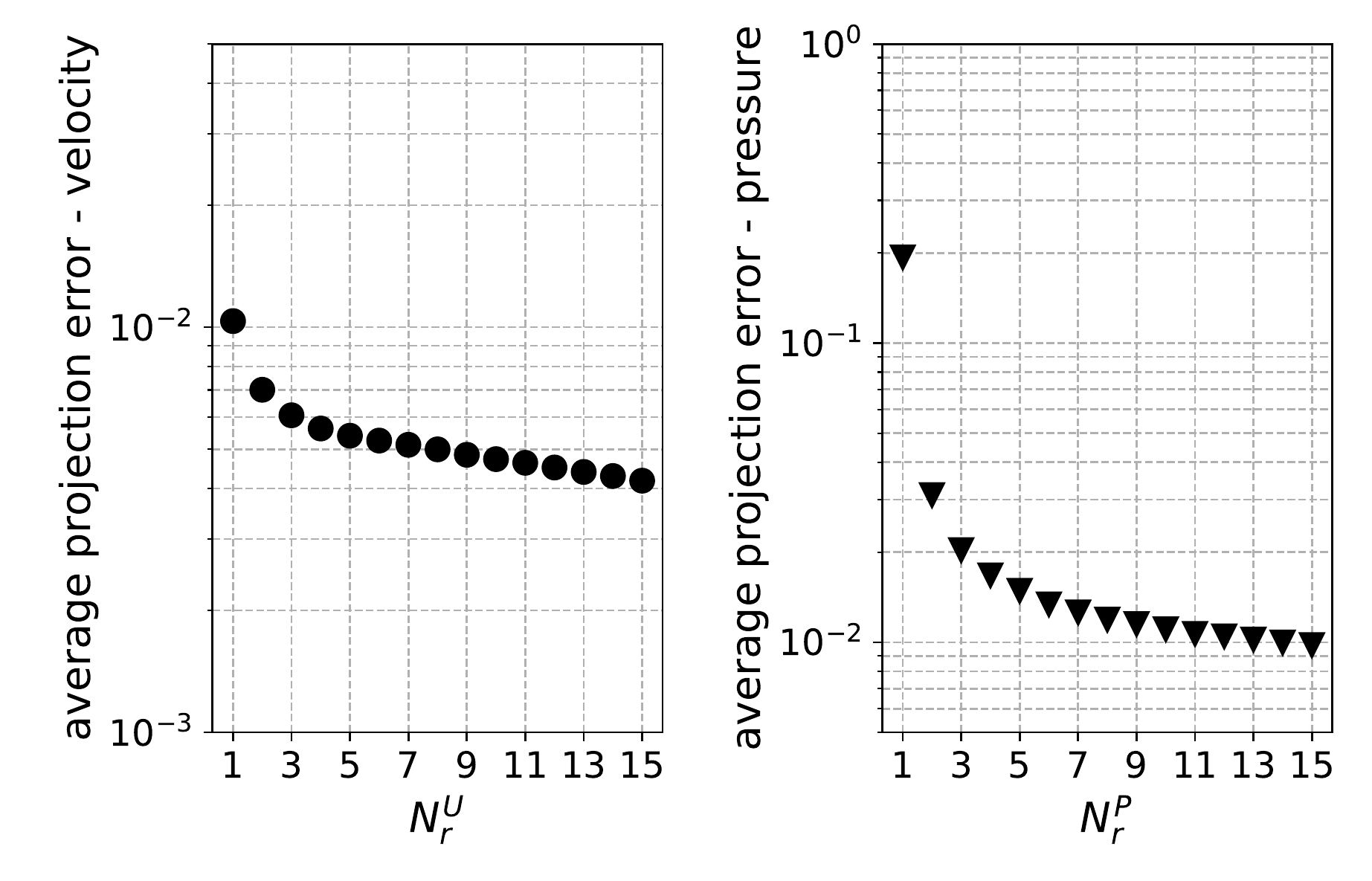}
\caption{The time-averaged relative basis projection error per number of modes for the open pipe flow test: (left) velocity; (right) pressure.}
\label{fig:ave_basis_proj}
\end{figure}

The coupled RELAP5/ROM model is tested and compared with the coupled RELAP5/CFD model. Snapshots are collected every 0.1 s for a pressure drop of 0.20 bar over the open pipe. Figure~\ref{fig:ave_basis_proj} shows the time-averaged basis projection error. The basis projection error is the relative error of the projected fields and the snapshots (Equation~\ref{eq:l2_projection}). The time-averaged projection error of velocity is of the order $\mathcal{O}$(\num{e-2}) and for pressure of the order $\mathcal{O}$(\num{e-1}). Only a few modes are needed to represent the flow solution due to the rapid decay of the error with increasing number of modes. Based on this, ten velocity and ten pressure modes are used to construct the reduced bases for this and all other test cases. The same number of eddy viscosity modes are used. In general, the number of POD modes should be determined based on the decay of eigenvalues.

\clearpage

\begin{figure}[h!]
\centering
\includegraphics[width=0.7\linewidth]{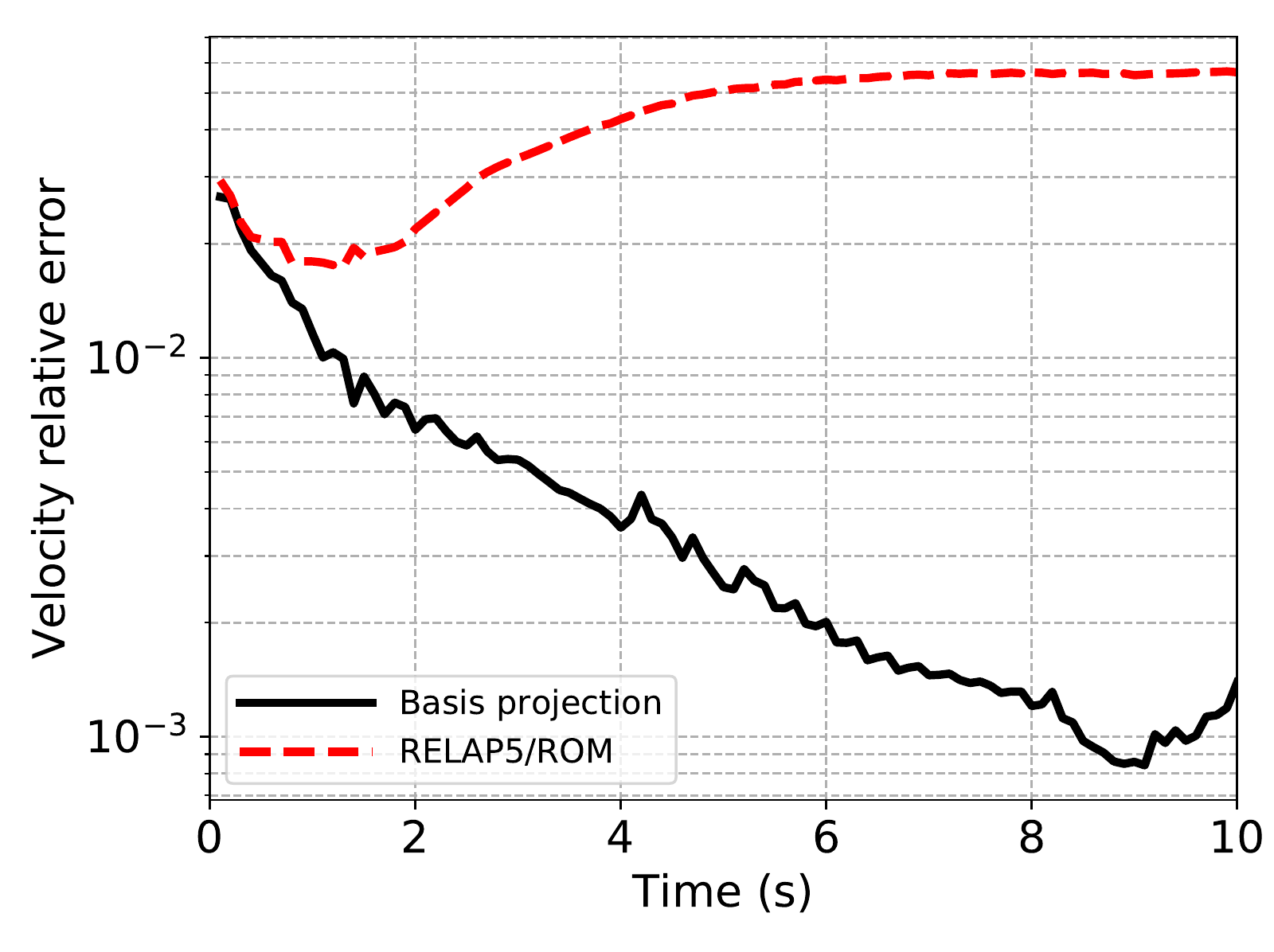}
\caption{Relative velocity error in the abrupt pressure difference transient ($\Delta P$ = 0.20 bar) for an open pipe flow configuration.$N_r = 10$ modes are used for the basis projection and ROM.}
\label{fig:res_open_rel_v}
\end{figure}

\begin{figure}[h!]
\centering
\includegraphics[width=0.7\linewidth]{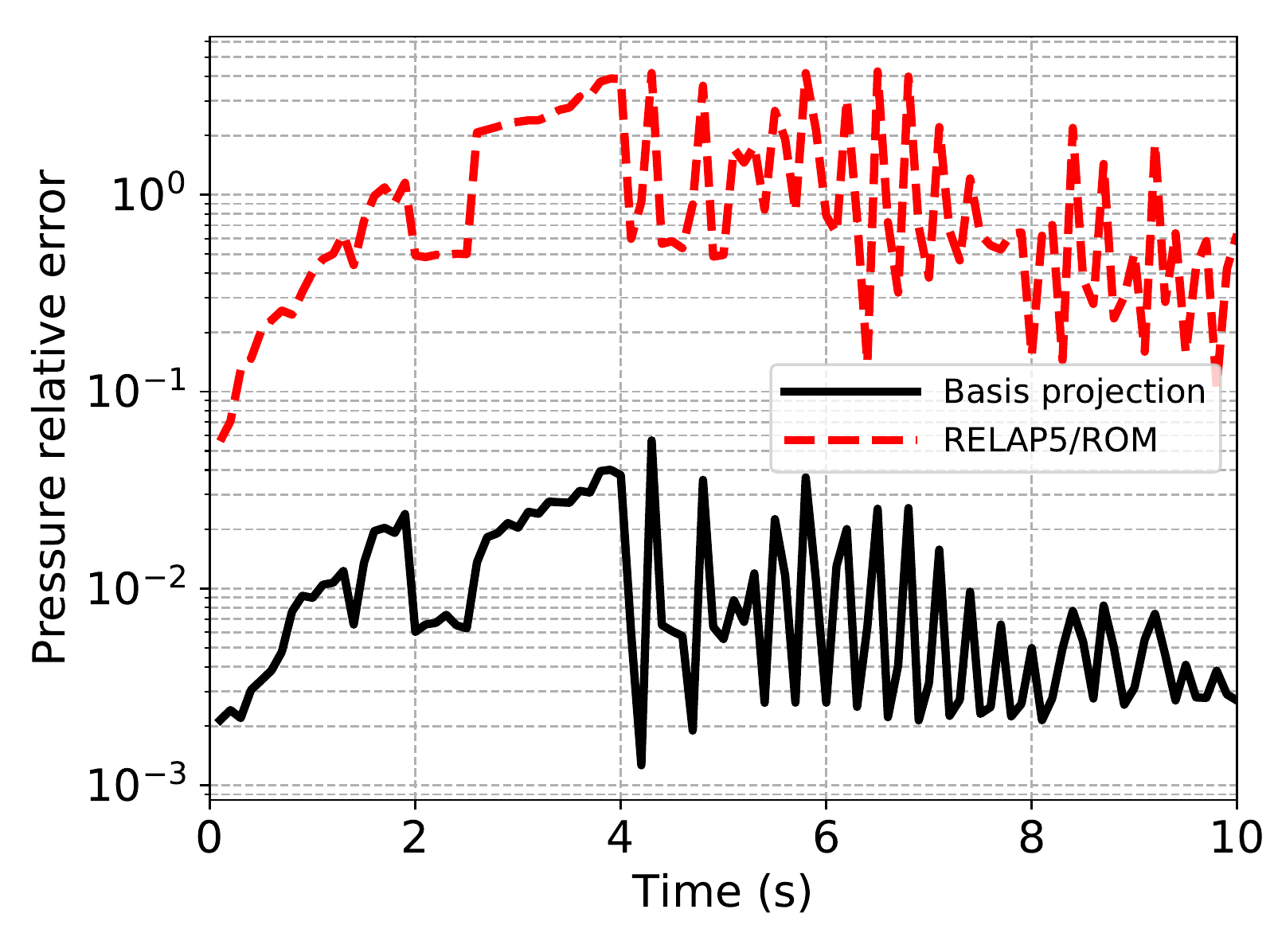}
\caption{Relative pressure error in the abrupt pressure difference transient ($\Delta P$ = 0.20 bar) for an open pipe flow configuration. $N_r = 10$ modes are used for the basis projection and ROM.}
\label{fig:res_open_rel_p}
\end{figure}

\begin{figure}[h!]
\centering
\includegraphics[width=0.7\linewidth]{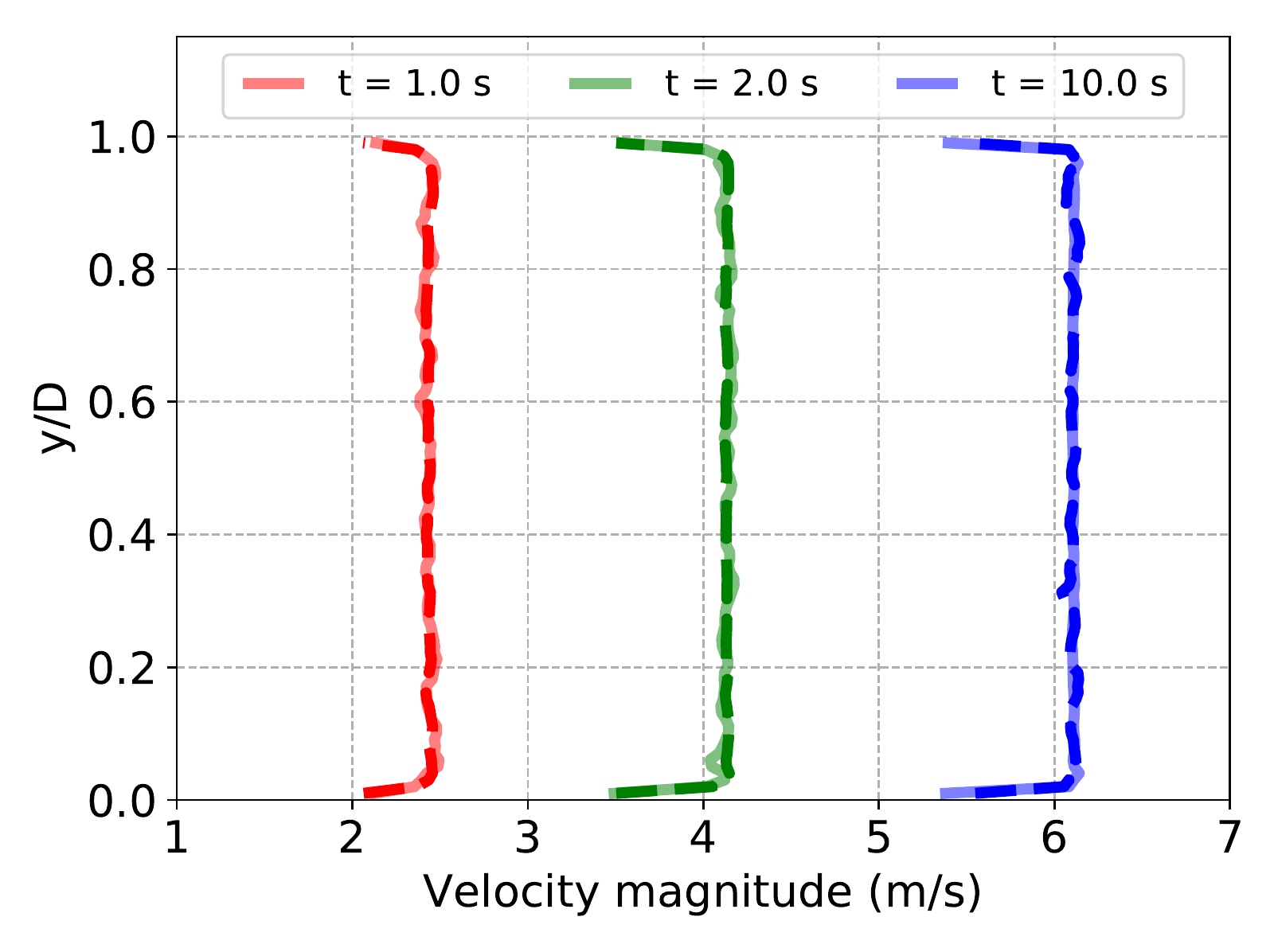}
\caption{Profiles of the velocity magnitude at $L_{CFD}/D$ = 0.5 downstream of the inlet of the CFD domain at $t$ = 1.0, 2.0 and 10.0 s of simulation time in the abrupt pressure difference transient ($\Delta P$ = 0.20 bar) for an open pipe flow configuration. Solid lines: RELAP5/CFD; Dashed lines: RELAP5/ROM.}
\label{fig:profiles_open_vel}
\end{figure}

Furthermore, the time evolution of the relative error (Equation~\ref{eq:l2_projection}) between the reconstructed fields and the RELAP5/CFD results is plotted in Figures~\ref{fig:res_open_rel_v} and~\ref{fig:res_open_rel_p}, which are compared with the basis projection error. The relative velocity error increases over time at the beginning of the simulation, while the basis projection error decreases over the whole simulation time. The pressure relative error is about two orders higher than the projection error. As velocity and pressure are coupled with the Pressure Poisson Equation, the velocity results are affected by the pressure results. Nevertheless, the relative velocity error stabilizes at about 5 \%. We also compare the profiles of the velocity magnitude in the CFD domain obtained with the RELAP5/ROM coupled model with the profiles obtained with the RELAP5/CFD model at $L_{CFD}/D$ = 0.5 at $t$ = 1.0, 2.0 and 10.0 s of simulation time in Figure~\ref{fig:profiles_open_vel}. Even though the relative L$^2$ error is of the order $\mathcal{O}$(\num{e-2}), the velocity profiles visually overlap. Therefore, the velocity results are considered reliable for the application studied.


\clearpage

\begin{figure}[h!]
\centering

\includegraphics[width=0.7\linewidth]{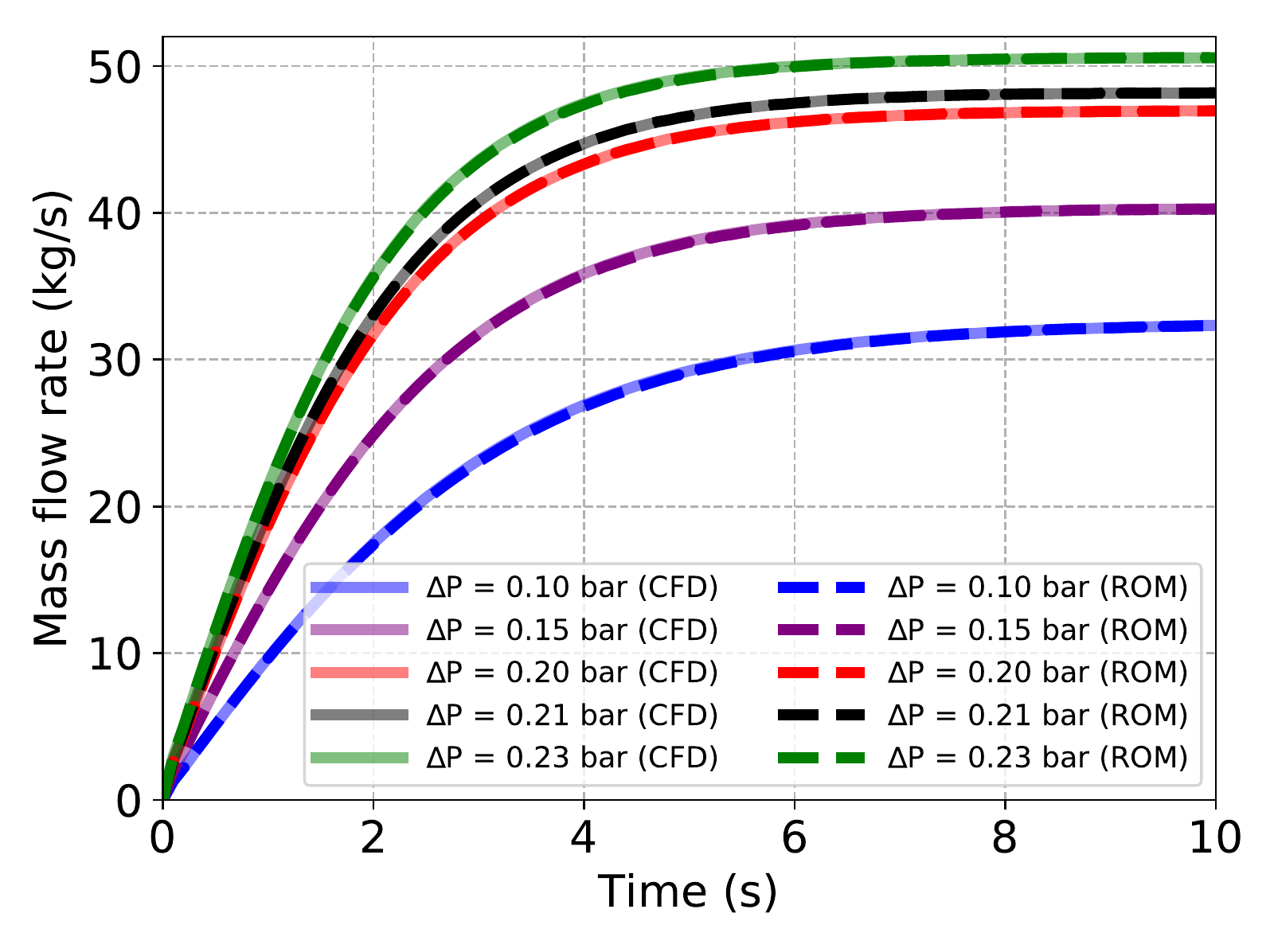}
\caption{Time evolution of the mass flow rate through the pipe at interface 2 in the abrupt pressure difference transient of different pressure drops for an open pipe flow configuration. $N_r = 10$ modes are used for the ROM.}
\label{fig:res_open_para_phi}
\end{figure}

\begin{figure}[h!]
\centering
\includegraphics[width=0.7\linewidth]{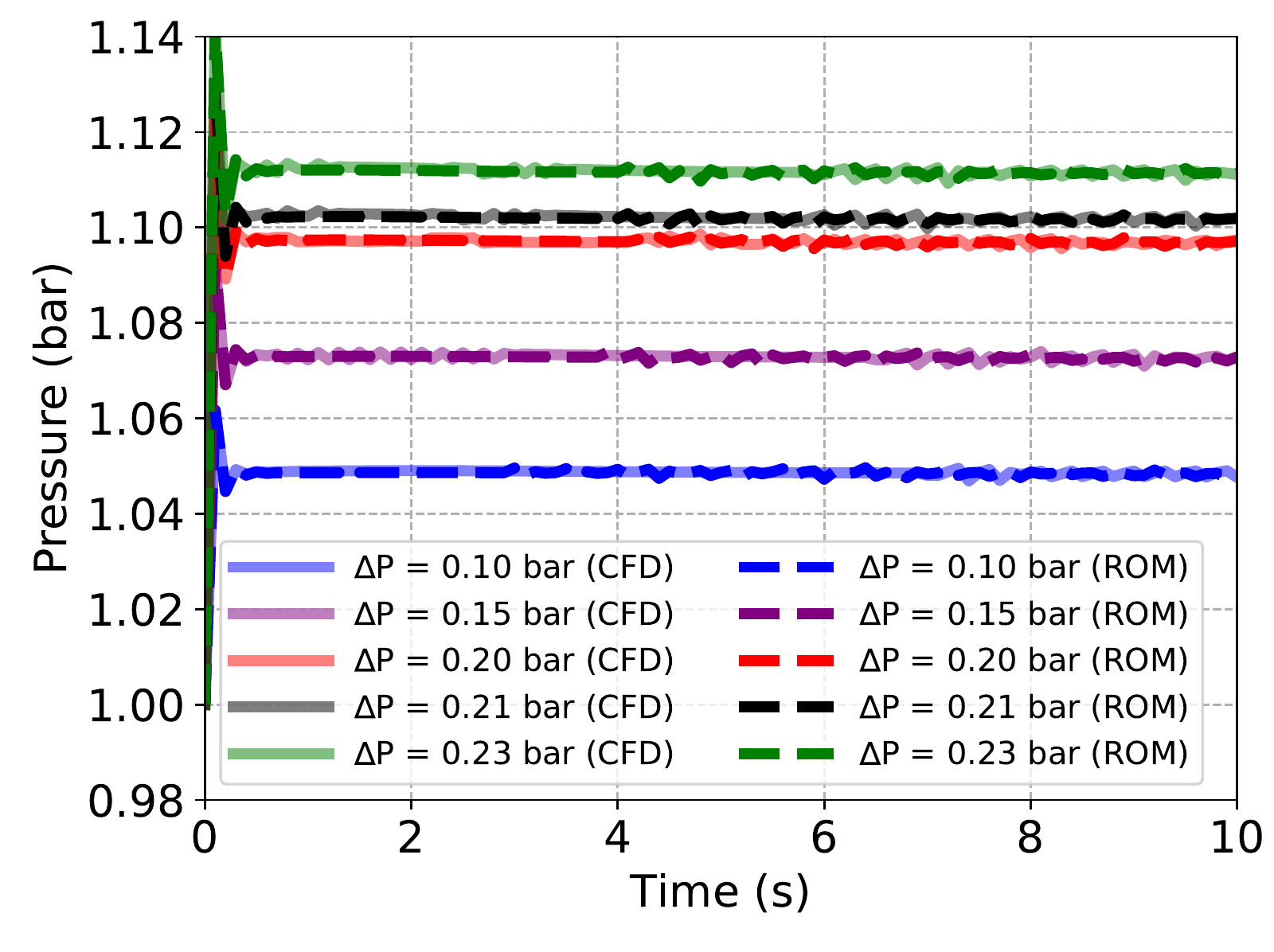}
\caption{Time evolution of the pressure at interface 2 in the abrupt pressure difference transient of different pressure drops for an open pipe flow configuration. $N_r = 10$ modes are used for the ROM.}
\label{fig:res_open_para_p}
\end{figure}

RELAP5/ROM simulations are performed for pressure drops of 0.10, 0.15, 0.20, 0.21 and 0.23 bar over the whole pipe. Figures~\ref{fig:res_open_para_phi} and~\ref{fig:res_open_para_p} show the time evolution of the mass flow rate at interface $\Gamma_2$ and the area-averaged pressure at the same interface, respectively. The RELAP5/ROM results visually overlap with the RELAP5/CFD results. Therefore, the ROM accurately predicts the CFD results for the entire parameter range even though the ROM is only constructed using the case with $\Delta P$ = 0.20 bar. However, the ROM is only valid in a range around the parameter used for the training. As information about the flow for lower pressure drops is contained in the snapshots, the ROM can even be used to simulate a pressure drop of 0.10 bar. However, when increasing the pressure drop more than 0.23 bar, the results become unphysical as the flow characteristics are not contained in the POD modes.  

The coupled RELAP5/CFD simulation of 10 seconds of simulation time takes about 2.8$\cdot$\num{e3} seconds for one parameter on one Intel\textsuperscript{\tiny\textregistered} Xeon\textsuperscript{\tiny\textregistered} core. On the other hand, one coupled RELAP5/ROM simulation takes about 7.0$\cdot$\num{e2} seconds on a single core. Therefore, the speed-up is about 4.0 times. The computational cost of the construction of the ROM (generating snapshots, calculating the POD modes and performing the Galerkin projection) is dominated by the time it takes to collect the snapshots. This cost is not taken into account in the calculation of the speed-up offered by the ROM itself.



\newpage

\subsection{Open pipe flow reversal test} \label{sec:res_reversal} 
\begin{figure}[h!]
\centering

\includegraphics[width=.7\linewidth]{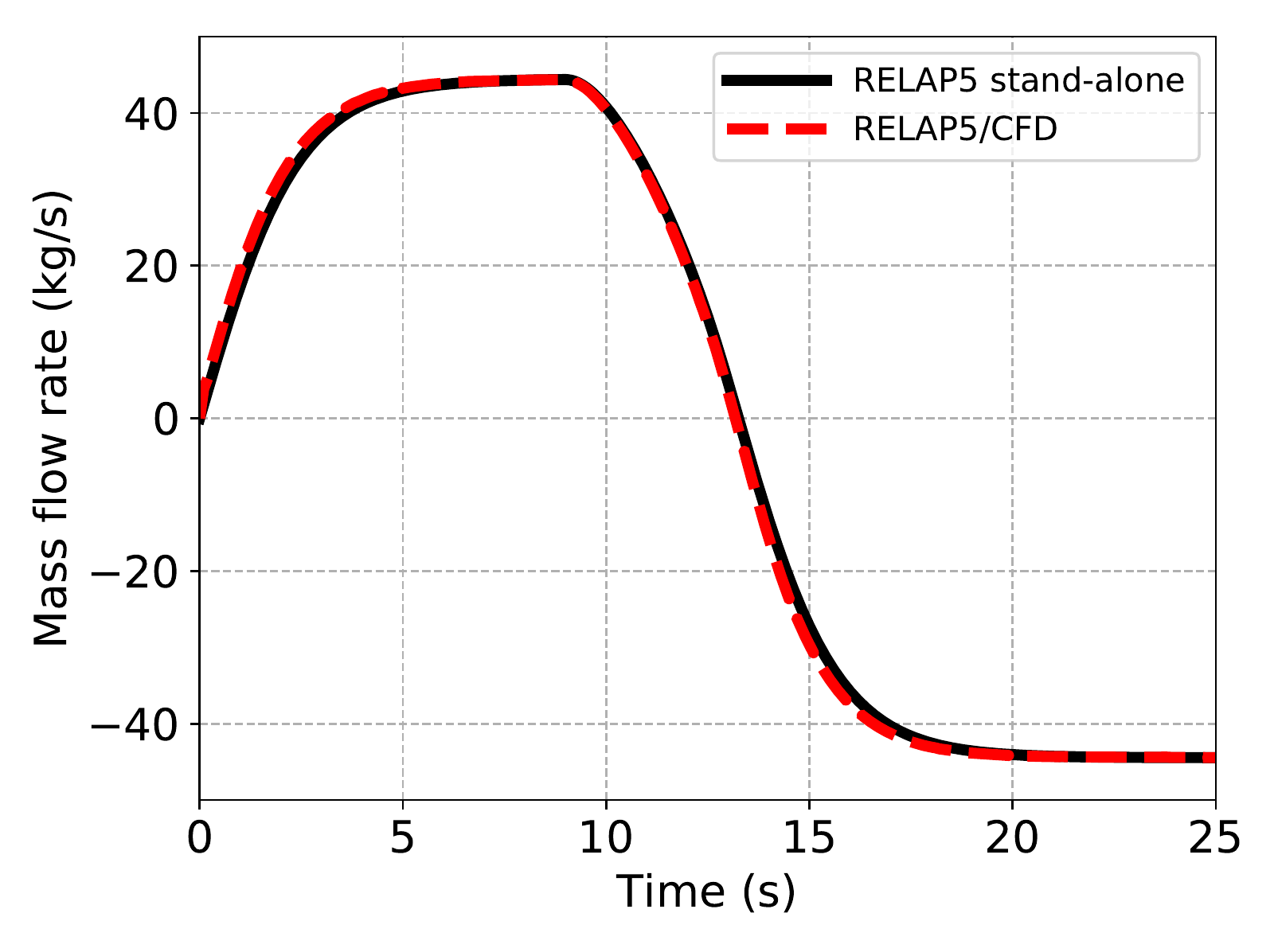}
\caption{Time evolution of the mass flow rate through the pipe at interface 2 in the abrupt forward and reverse pressure difference transient for a pressure drop of 0.20 bar over the open pipe.}
\label{fig:res_reversal_phi}
\end{figure}

\begin{figure}[h!]
\centering
\includegraphics[width=0.7\linewidth]{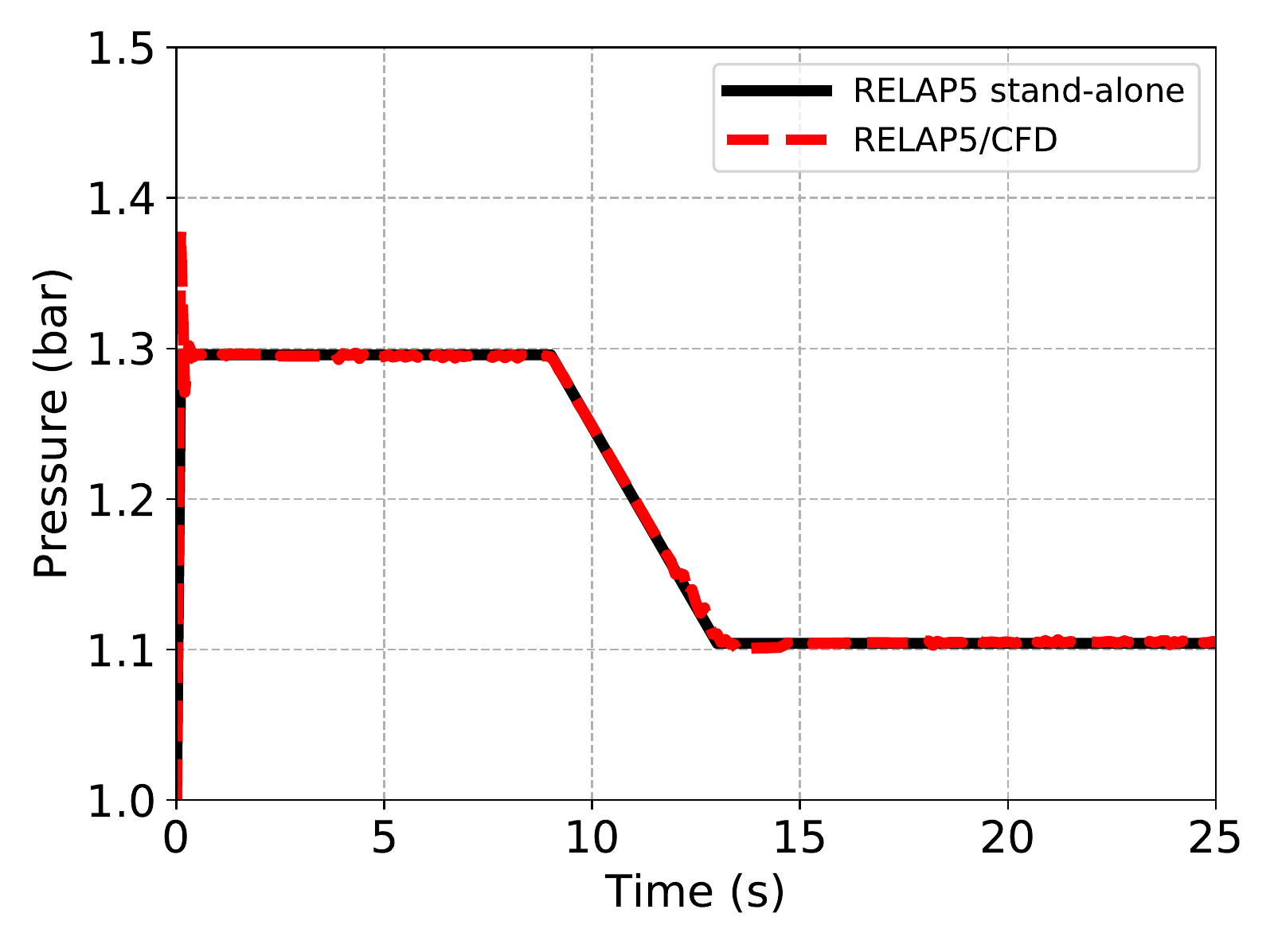}

\caption{Time evolution of the pressure at interface 2 in the abrupt forward and reverse pressure difference transient for a pressure drop of 0.20 bar over the open pipe.}
\label{fig:res_reversal_p}
\end{figure}

Coupled simulations are performed for the open pipe flow reversal test case. First, the coupled RELAP5/CFD results for a pressure drop of 0.20 bar are compared with RELAP5 stand-alone simulation as shown in Figure~\ref{fig:res_reversal_phi} for the mass flow rate and Figure~\ref{fig:res_reversal_p} for pressure at interface $\Gamma_2$. Similar to previous test case, oscillations are present in the pressure. Especially at the beginning of the simulation, there is a spike in pressure obtained with the coupled system compared to the RELAP5 stand alone simulation. Oscillations also occur when the drop in mass flow rate (between 12 and 13 seconds of simulations time) is the steepest. These do however disappear as the simulation time proceeds. 



\begin{figure}[h!]
\centering
\includegraphics[width=0.7\linewidth]{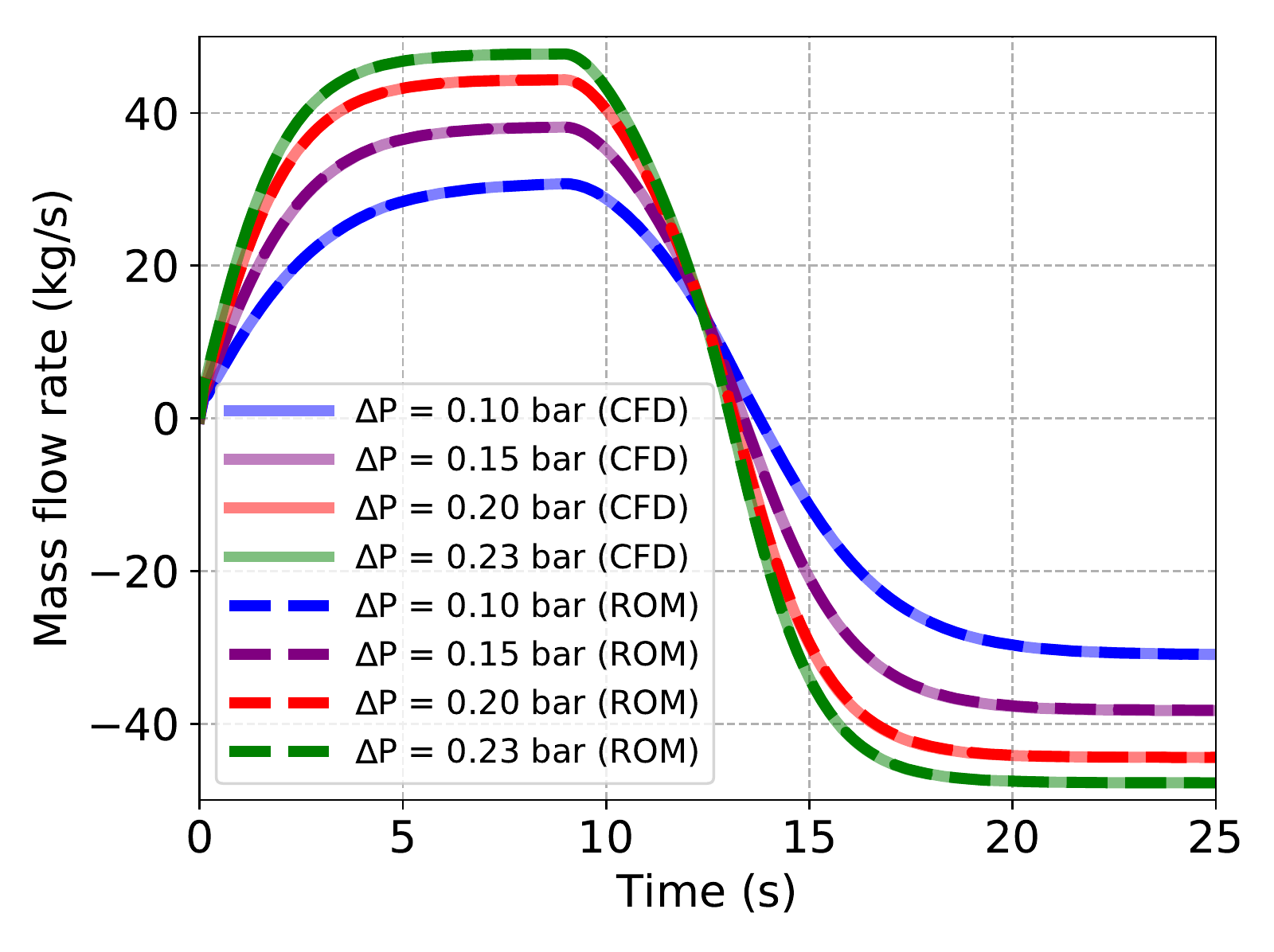}
\caption{Time evolution of the mass flow rate through the pipe at interface 2 in the abrupt forward and reverse pressure difference transient for different pressure drops for an open pipe flow configuration. $N_r = 10$ modes are used for the ROM.}
\label{fig:res_reversal_para_phi}
\end{figure}

\begin{figure}[h!]
\centering
\includegraphics[width=0.7\linewidth]{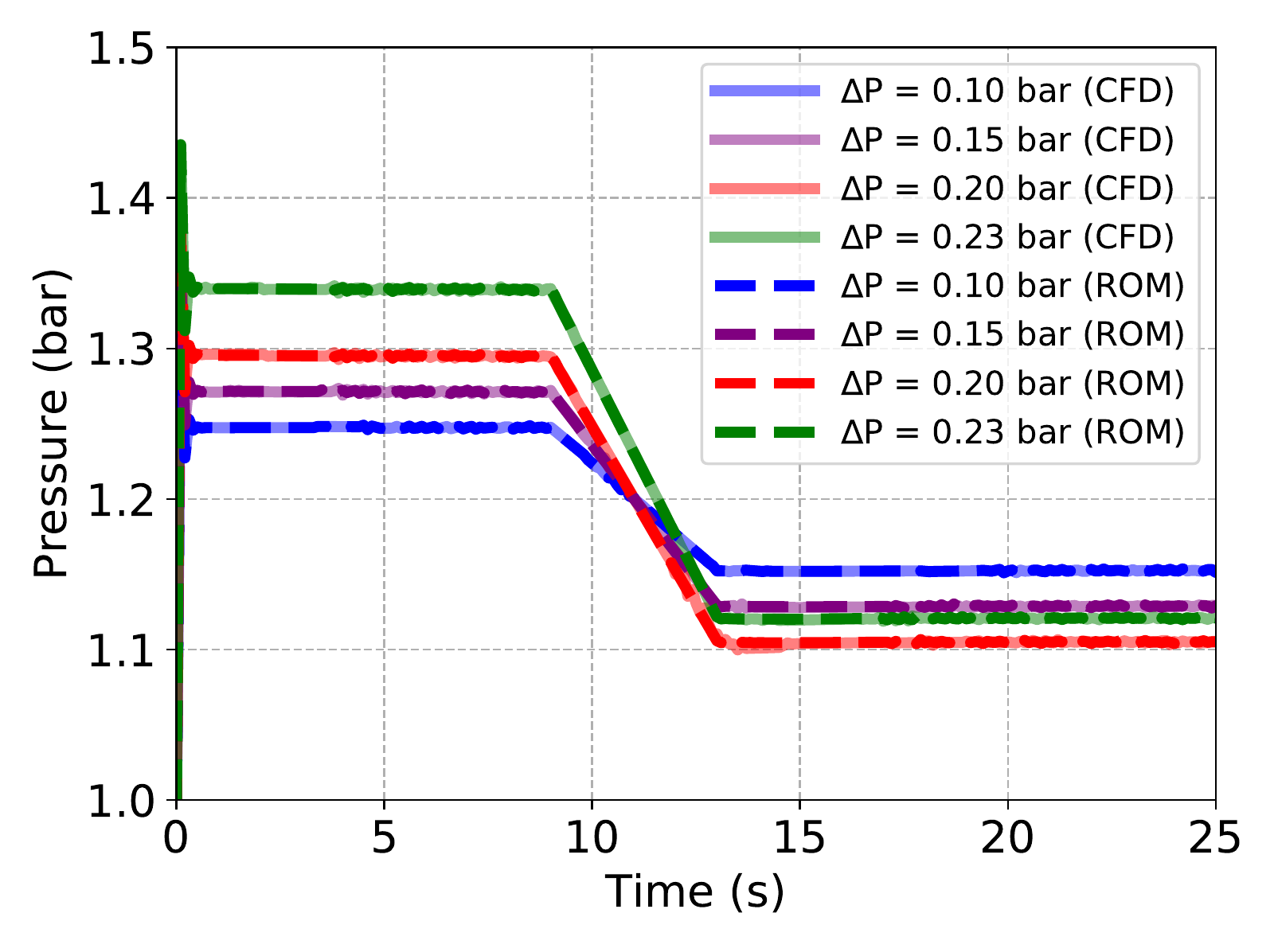}
\caption{Time evolution of the pressure at interface 2 in the abrupt forward and reverse pressure difference transient for different pressure drops for an open pipe flow configuration. $N_r = 10$ modes are used for the ROM.}
\label{fig:res_reversal_para_p}
\end{figure}

\newpage
Furthermore, the coupled RELAP5/ROM model is tested for several new values of the pressure drop over the pipe and compared to the coupled RELAP5/CFD results in Figures~\ref{fig:res_reversal_para_phi} and~\ref{fig:res_reversal_para_p}. The figures show that the ROM is capable of predicting the FOM results within the tested range of parameter values. 

%

The coupled RELAP5/CFD simulation takes about 7.4$\cdot$\num{e3} seconds for one parameter on a single core. On the other hand, one coupled RELAP5/ROM simulation takes about 1.6$\cdot$\num{e3} seconds on a single core. Therefore, the speed-up is about 4.6 times. 

\newpage

\subsubsection{Long time integration}
\begin{figure}[h!]
\centering
\includegraphics[width=0.7\linewidth]{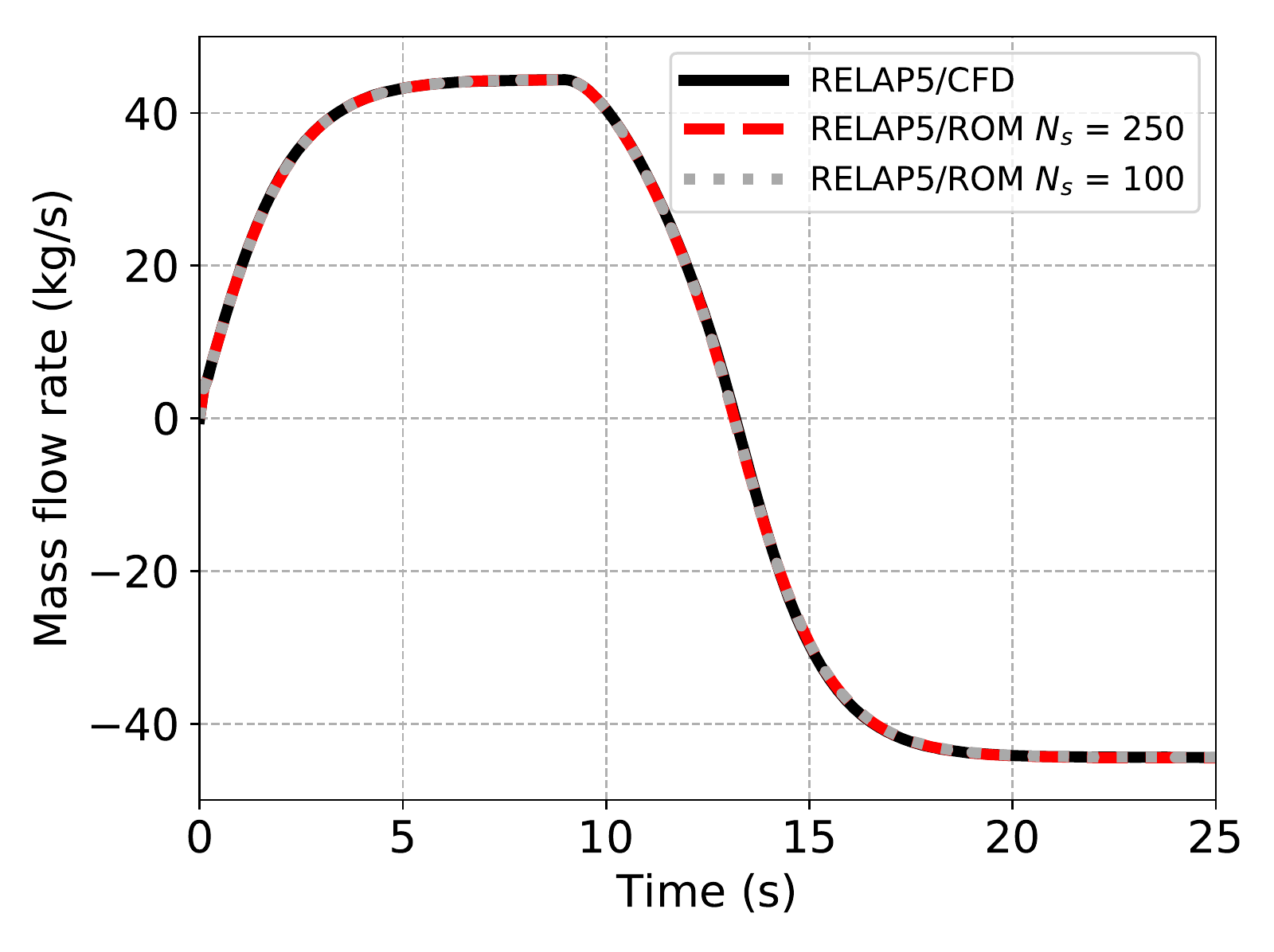}
\caption{Time evolution of the mass flow rate through the pipe at interface 2 in the long term integration test ($\Delta P$ = 0.20 bar) for an open pipe flow configuration. $N_r = 10$ modes are used for the ROM.}
\label{fig:res_reversal_lti_phi}
\end{figure}

\begin{figure}[h!]
\centering
\includegraphics[width=0.7\linewidth]{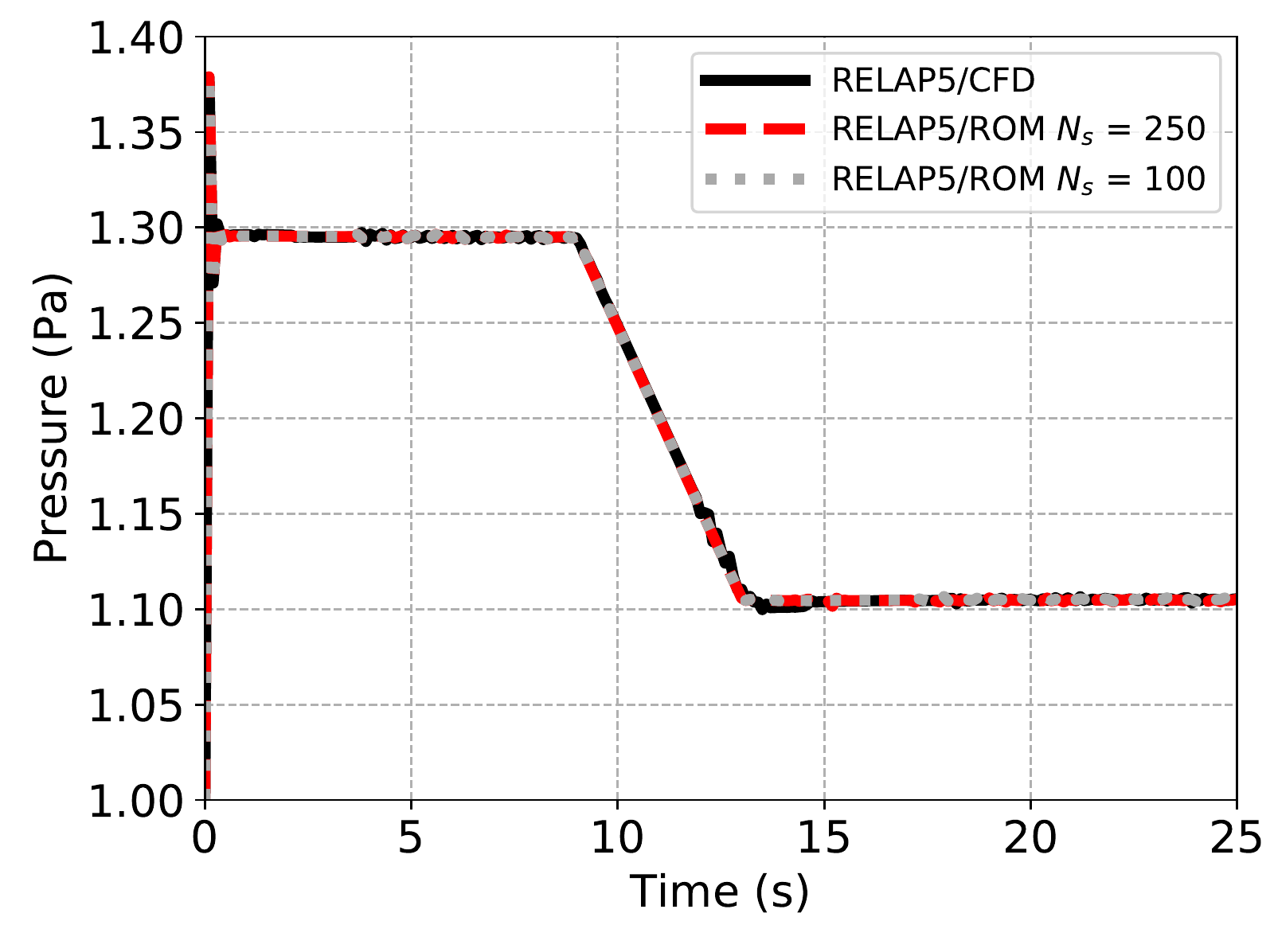}
\caption{Time evolution of the pressure at interface 2 in the long term integration test ($\Delta P$ = 0.20 bar) for an open pipe flow configuration. $N_r = 10$ modes are used for the ROM.}
\label{fig:res_reversal_lti_p}
\end{figure}

The coupled RELAP5/ROM model is also tested for long-term integration. Snapshots are collected using the RELAP5/CFD model for the reverse flow test case up to 10 seconds of simulation time. Thus, in total 100 snapshots are collected for the reduced basis construction. A RELAP5/ROM simulation is then performed up to 25 seconds of simulation time. The results for a pressure drop of 0.20 bar are shown in Figures~\ref{fig:res_reversal_lti_phi} and~\ref{fig:res_reversal_lti_p}, which are compared with the previous RELAP5/ROM model of which the reduced basis is constructed with all 250 snapshots. The RELAP5/ROM model fully predicts the behavior of the RELAP5/CFD model at coupling interface 2 as the results for the time evolution of the mass flow rate and pressure correspond to those of the RELAP5/ROM model based on 250 snapshots. It is important to note that this is true at the coupling interfaces.

%

\newpage

\begin{figure}[h!]
\centering
\includegraphics[width=0.7\linewidth]{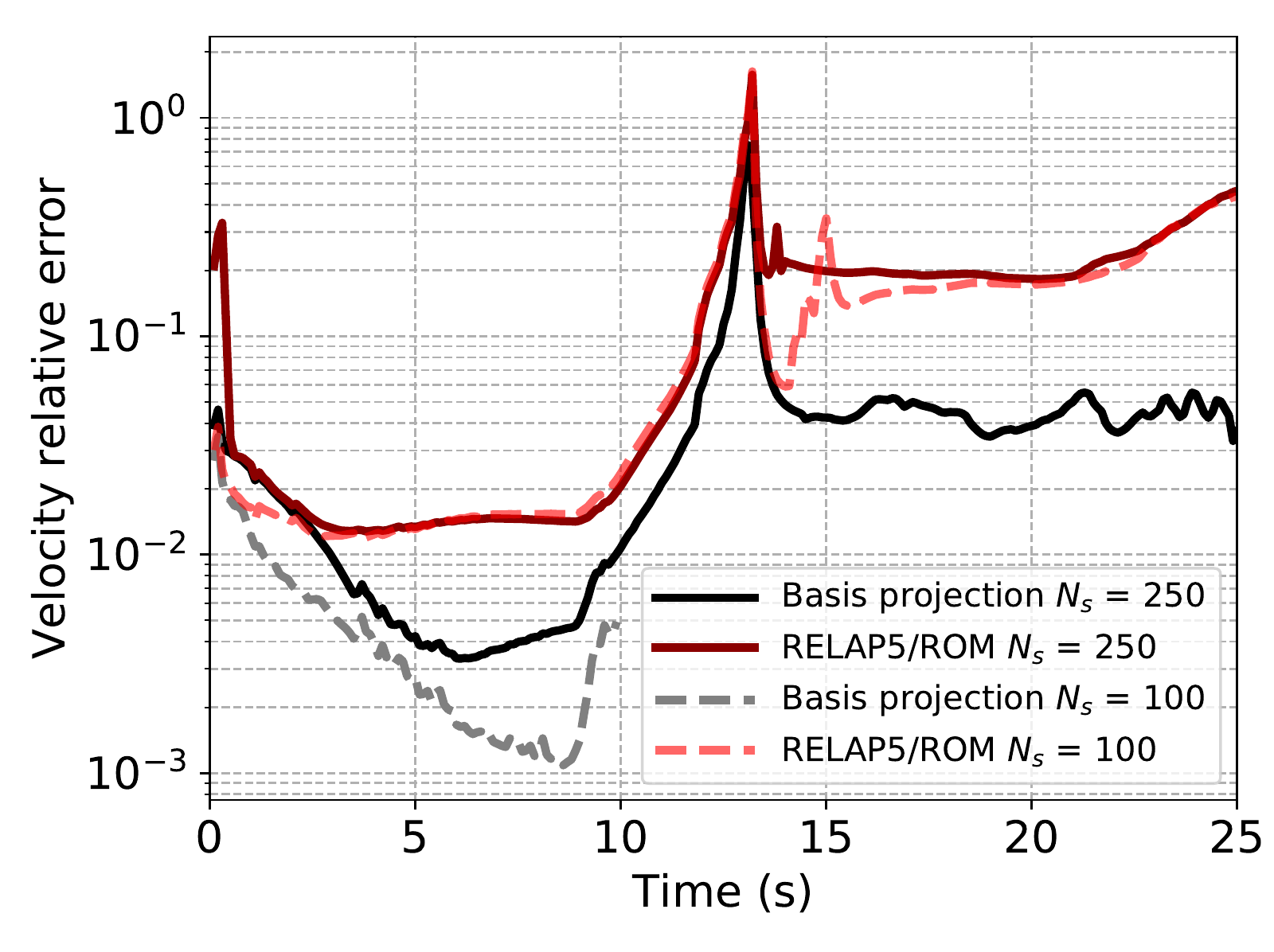}
\caption{Relative velocity error at interface 2 in the long term integration test ($\Delta P$ = 0.20 bar) for an open pipe flow configuration. $N_r = 10$ modes are used for the basis projection and ROM.}
\label{fig:res_reversal_rel_v}
\end{figure}

\begin{figure}[h!]
\centering
\includegraphics[width=0.7\linewidth]{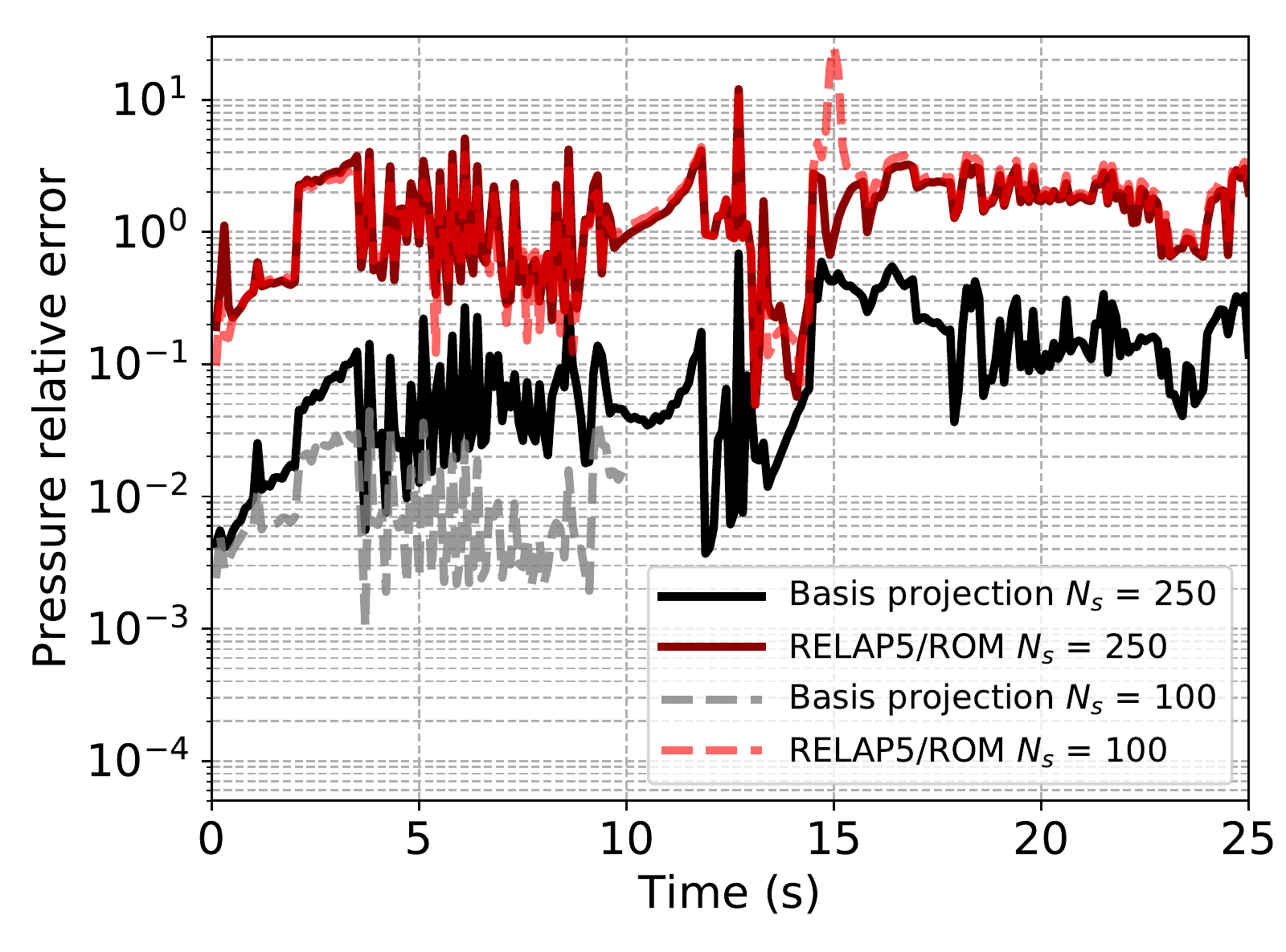}
\caption{Relative pressure error at interface 2 in the long term integration test ($\Delta P$ = 0.20 bar) for an open pipe flow configuration. $N_r = 10$ modes are used for the basis projection and ROM.}
\label{fig:res_reversal_rel_p}
\end{figure}

The relative error of the solution in the CFD domain for the coupled RELAP5/ROM and coupled RELAP5/ROM on the long term integration are plotted in Figures~\ref{fig:res_reversal_rel_v} and~\ref{fig:res_reversal_rel_p} for the velocity and pressure, respectively. For all models, the relative error spikes at about 13 seconds of simulation time. Around this time the decrease in mass flow rate is the steepest. 

Even though only the first 100 snapshots, obtained till 10 seconds of simulation time, are included in the reduced order model in case of long time integration, the RELAP5/ROM model is capable of reproducing the RELAP5/CFD results. This means that the overall mass flow through the pipe is conserved. Also the relative pressure errors are of the same order.



\begin{figure}[h!]
\centering
\includegraphics[width=0.7\linewidth]{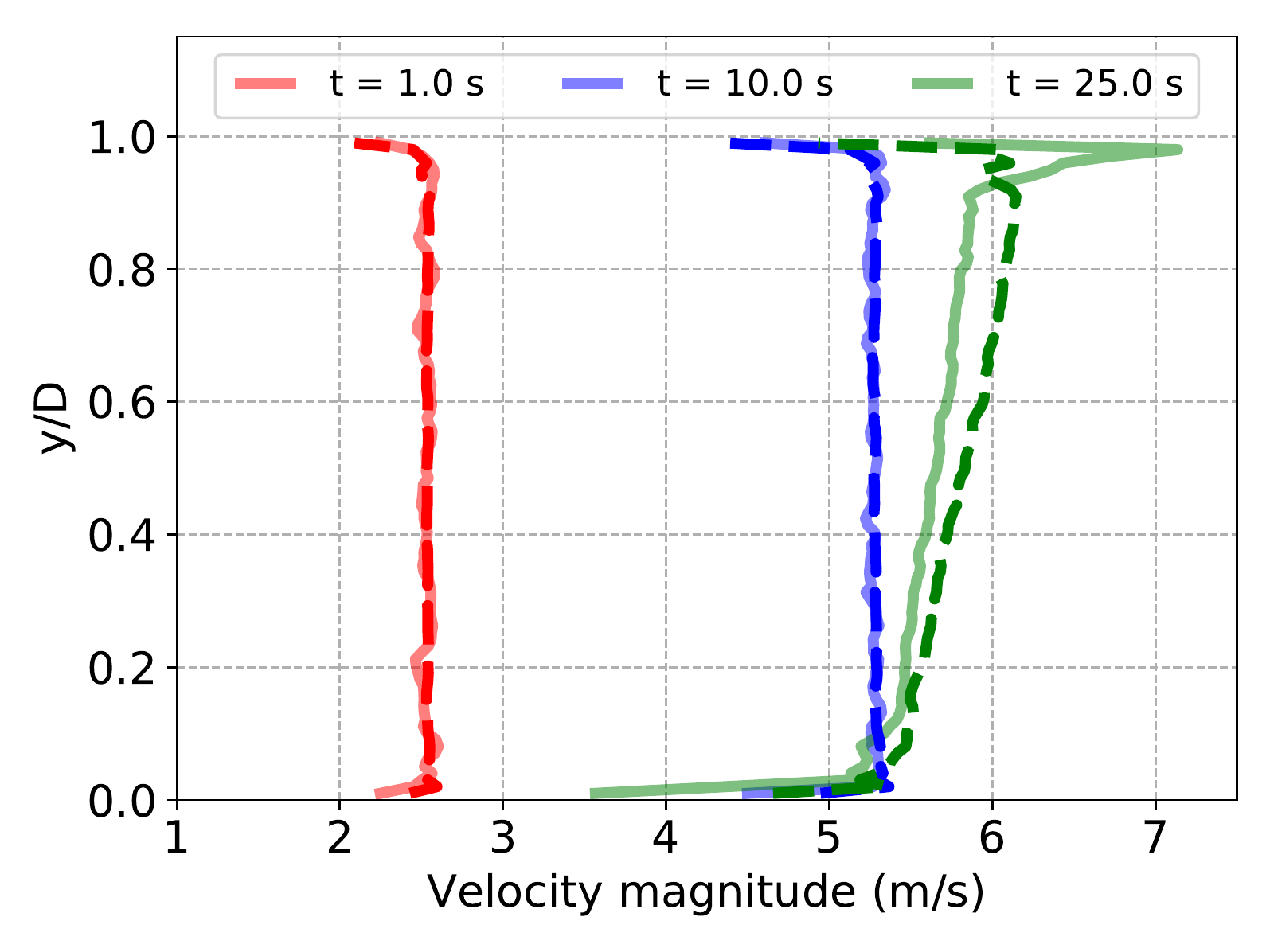}
\caption{Profiles of the velocity magnitude at $L_{CFD}/D$ = 0.5 downstream of the inlet of the CFD domain at $t$ = 1.0, 10.0 and 25.0 s of simulation time of the reverse pressure difference transient for a pressure drop of 0.20 bar over the open pipe. Solid lines: RELAP5/CFD; Dashed lines: RELAP5/ROM.}
\label{fig:profiles_reversal_vel}
\end{figure}

\newpage 
Figure~\ref{fig:profiles_reversal_vel} shows the profiles of the velocity magnitude obtained with the RELAP5/CFD and RELAP5/ROM trained with 250 snapshots at $L_{CFD}/D$ = 0.5 downstream of the inlet of the CFD domain at $t$ = 1.0, 10.0 and 25.0 s of simulation time. The results are visually overlapping, except at the final simulation time ($t$ = 25.0 s). This corresponds to the relative error plotted in Figure~\ref{fig:res_reversal_rel_v}, which is about one order higher at $t$ = 25.0 s compared to $t$ = 1.0 and 10.0 s of simulation time.

Finally, the RELAP5/ROM model accurately determines the mass flow rate and pressure at coupling interface 2 during the flow reversal test and does not exhibit instabilities even outside the time domain in which snapshots were collected.

\subsection{Closed pipe flow test} \label{sec:res_closed_pipe} 
\begin{figure}[h!]
\centering
\includegraphics[width=0.7\linewidth]{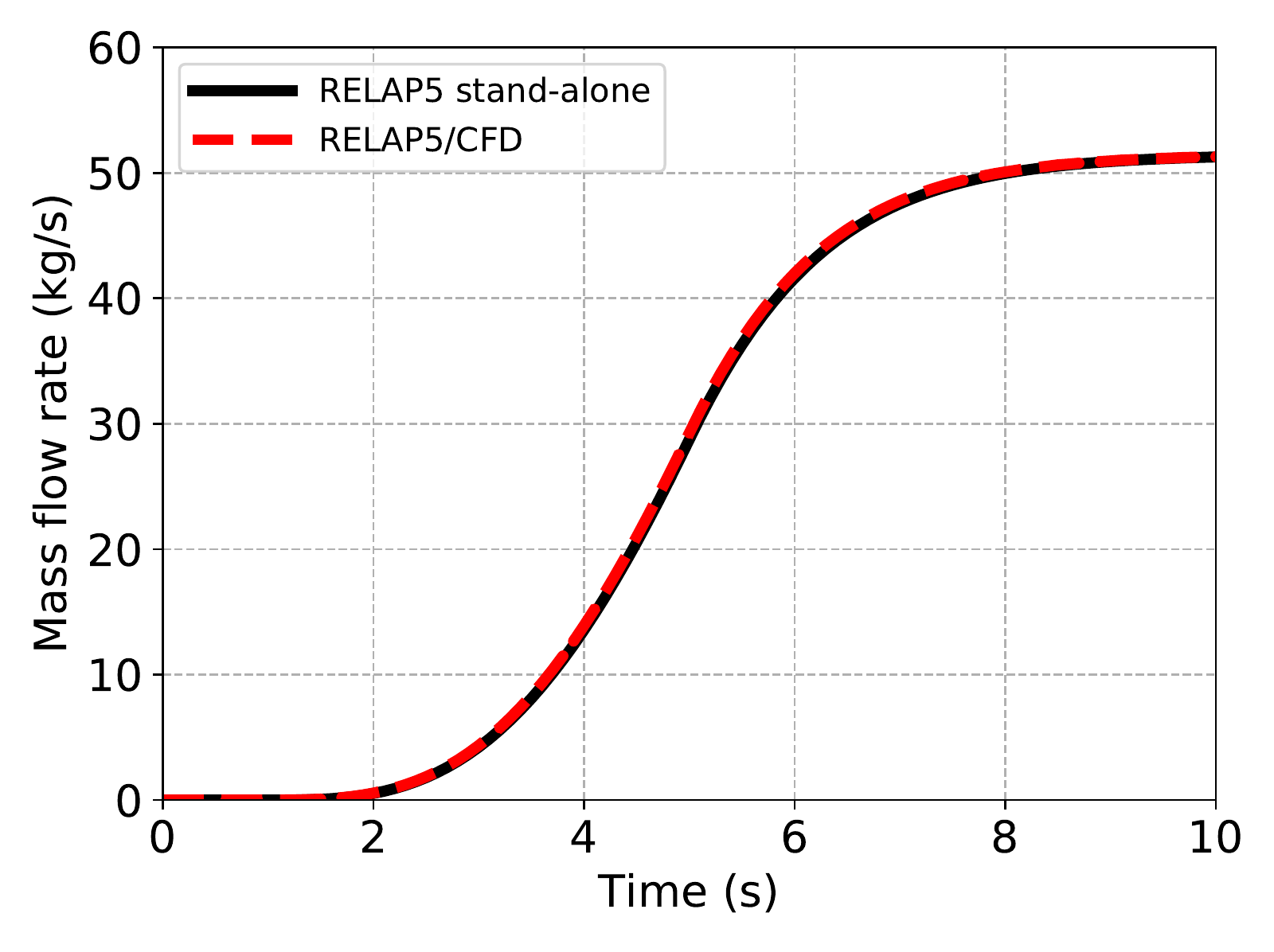}
\caption{Time evolution of the mass flow rate through the pipe at interface 2 for the closed pipe flow test with a maximum pump rotor rotational speed of 100 rad/s.}
\label{fig:res_closed_phi}
\end{figure}

\begin{figure}[h!]
\centering
\includegraphics[width=0.7\linewidth]{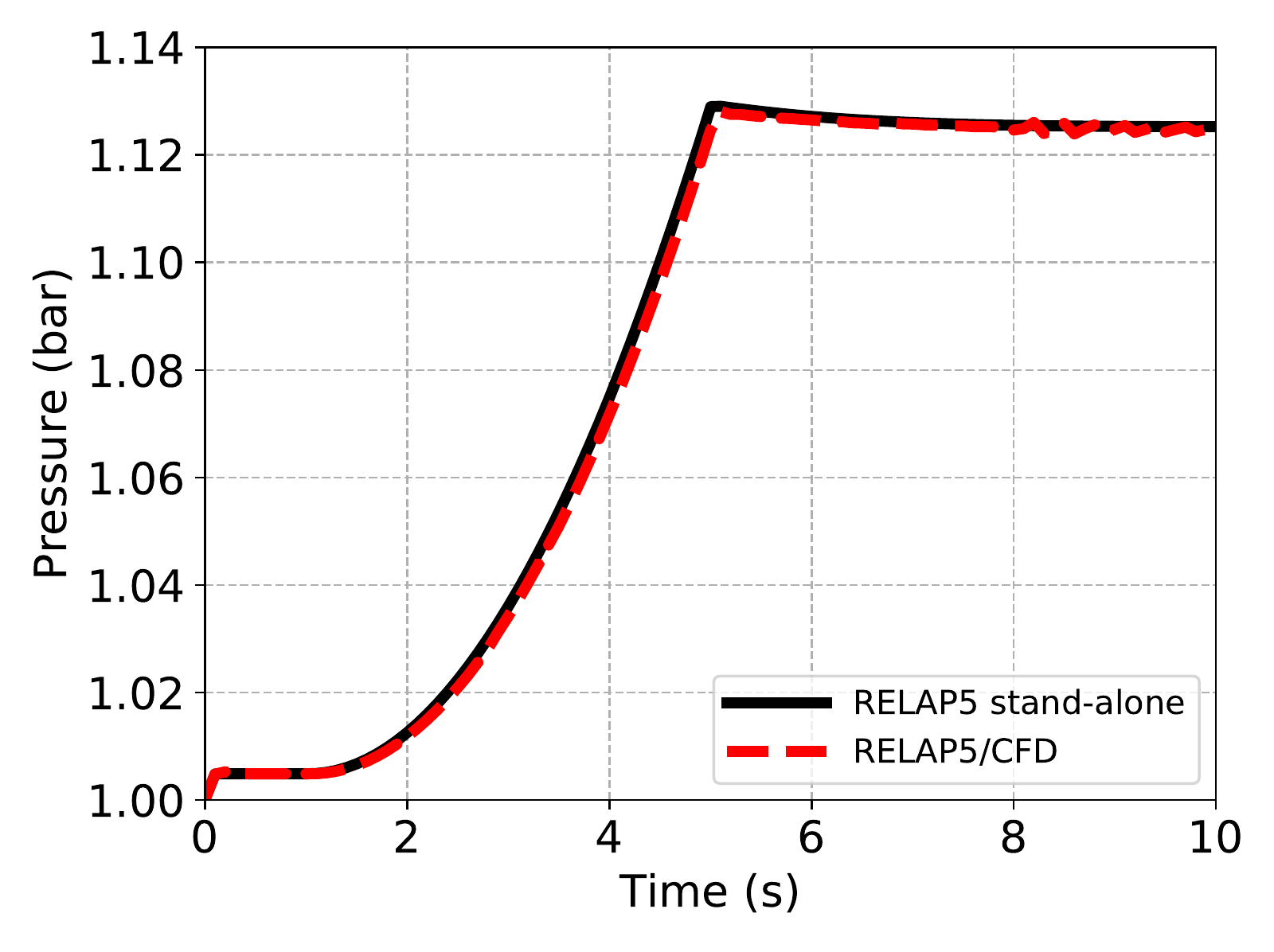}
\caption{Time evolution of the pressure  at interface 2 for the closed pipe flow test with a maximum pump rotor rotational speed of 100 rad/s.}
\label{fig:res_closed_p}
\end{figure}

The coupling methodology is also tested on the closed pipe flow test case. Snapshots are collected for a the maximum pump rotor rotational speed of 100 rad/s and the results for the mass flow rate and pressure at coupling interface 2 are compared with a RELAP5 stand alone simulation in Figures~\ref{fig:res_closed_phi} and~\ref{fig:res_closed_p}, respectively.

\clearpage

\begin{figure}[h!]
\centering
\includegraphics[width=0.7\linewidth]{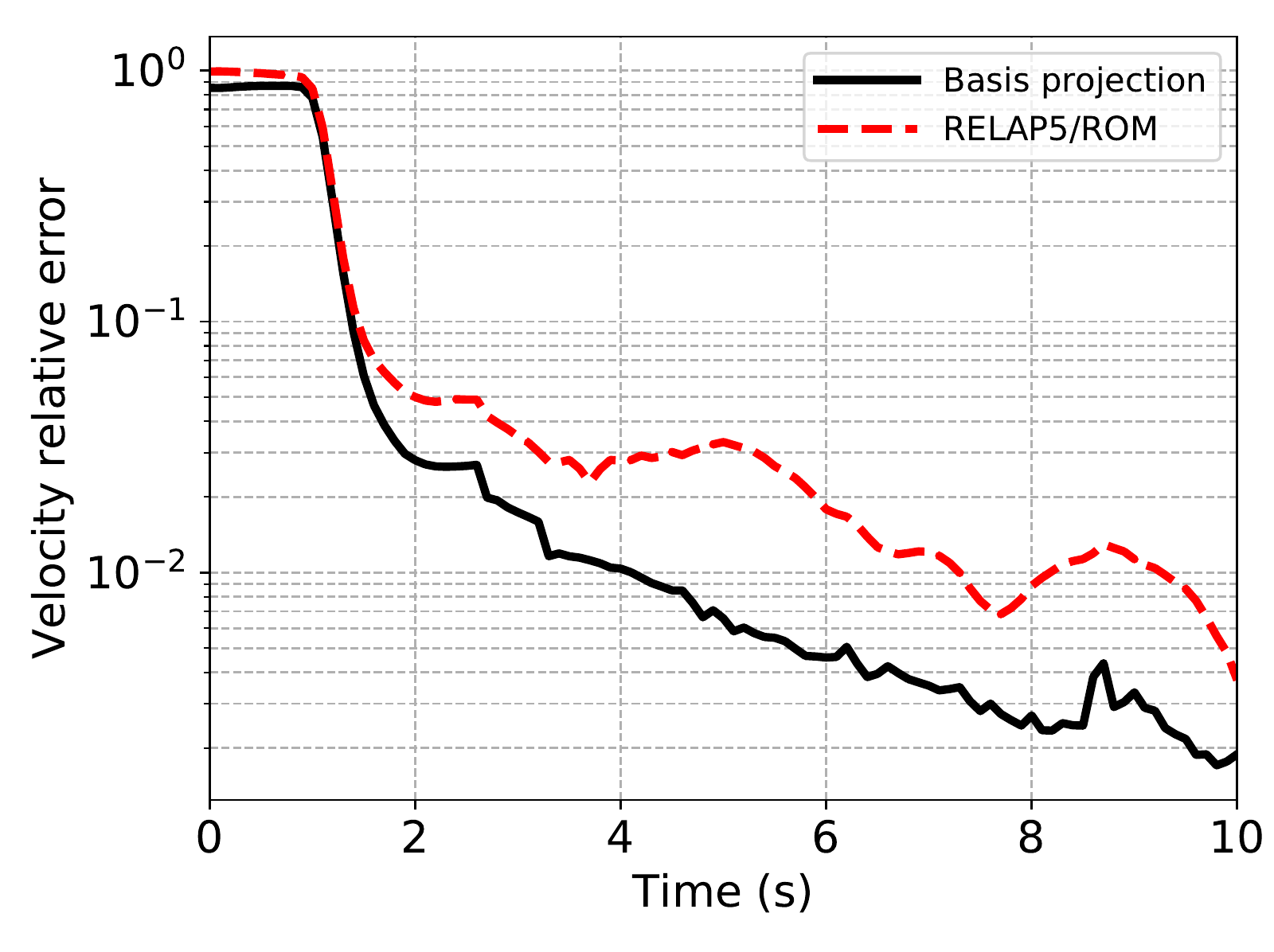}
\caption{Relative velocity error for the closed pipe flow test with a maximum pump rotor rotational speed of 100 rad/s. $N_r = 10$ modes are used for the basis projection and ROM.}
\label{fig:res_closed_rel_v}
\end{figure}

\begin{figure}[h!]
\centering
\includegraphics[width=0.7\linewidth]{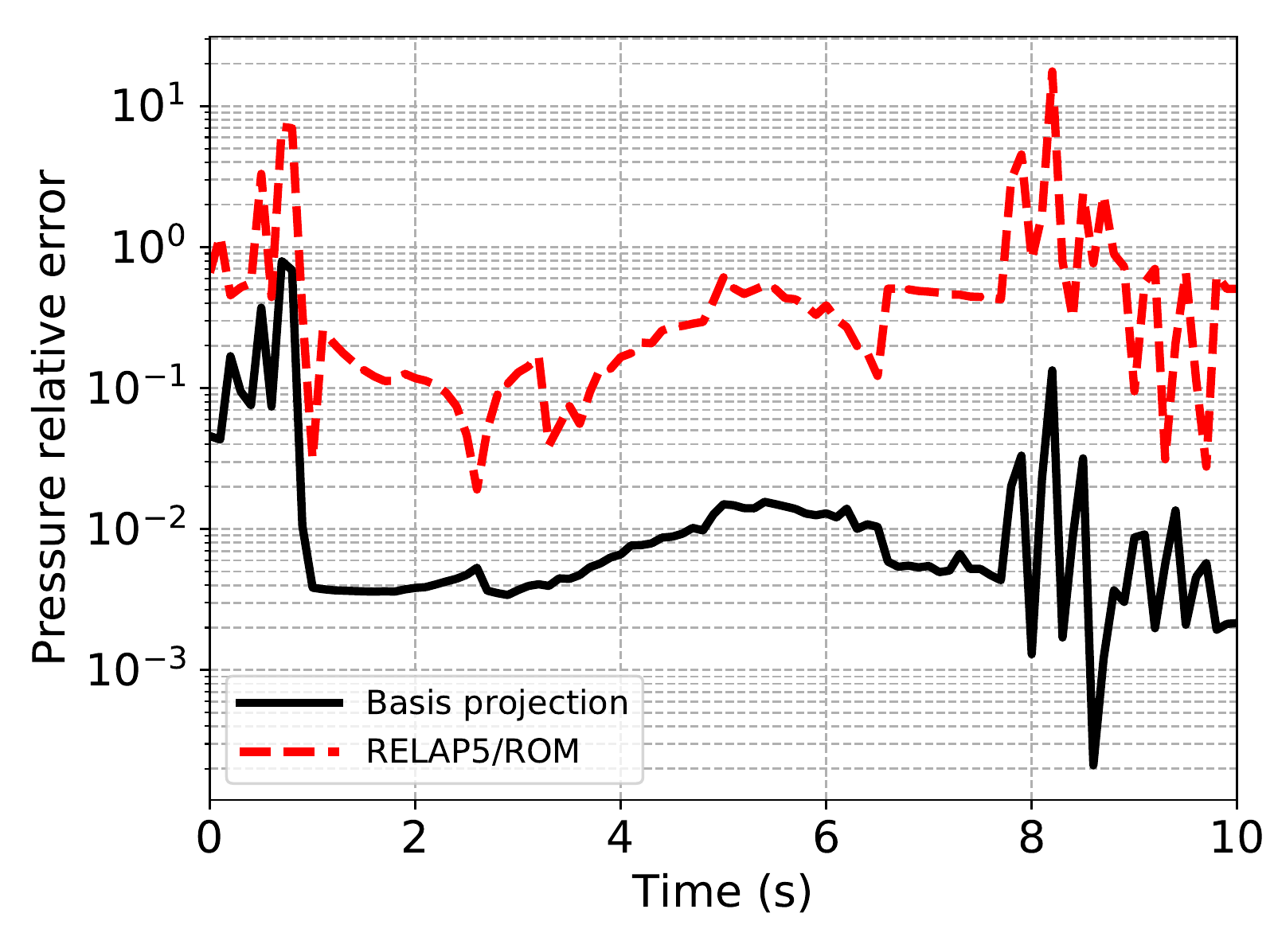}
\caption{Relative pressure error for the closed pipe flow test with a maximum pump rotor rotational speed of 100 rad/s. $N_r = 10$ modes are used for the basis projection and ROM.}
\label{fig:res_closed_rel_p}
\end{figure}

The relative velocity and pressure errors are plotted in Figures~\ref{fig:res_closed_rel_v} and~\ref{fig:res_closed_rel_p}, respectively. The velocity relative error is the largest at the beginning of the simulation. As the flow is initially at rest, a small error in the reconstructed flow field results in large relative error. As soon as the mass flow rate increases, the relative error drops. This also indicates that the closed pipe flow test case is numerically more stable than the open pipe flow test, where the relative velocity error increased as function of time as shown in Figure~\ref{fig:res_open_rel_v}. The relative pressure error is about two orders larger than the projection error.

\begin{figure}[h!]
\centering
\includegraphics[width=0.7\linewidth]{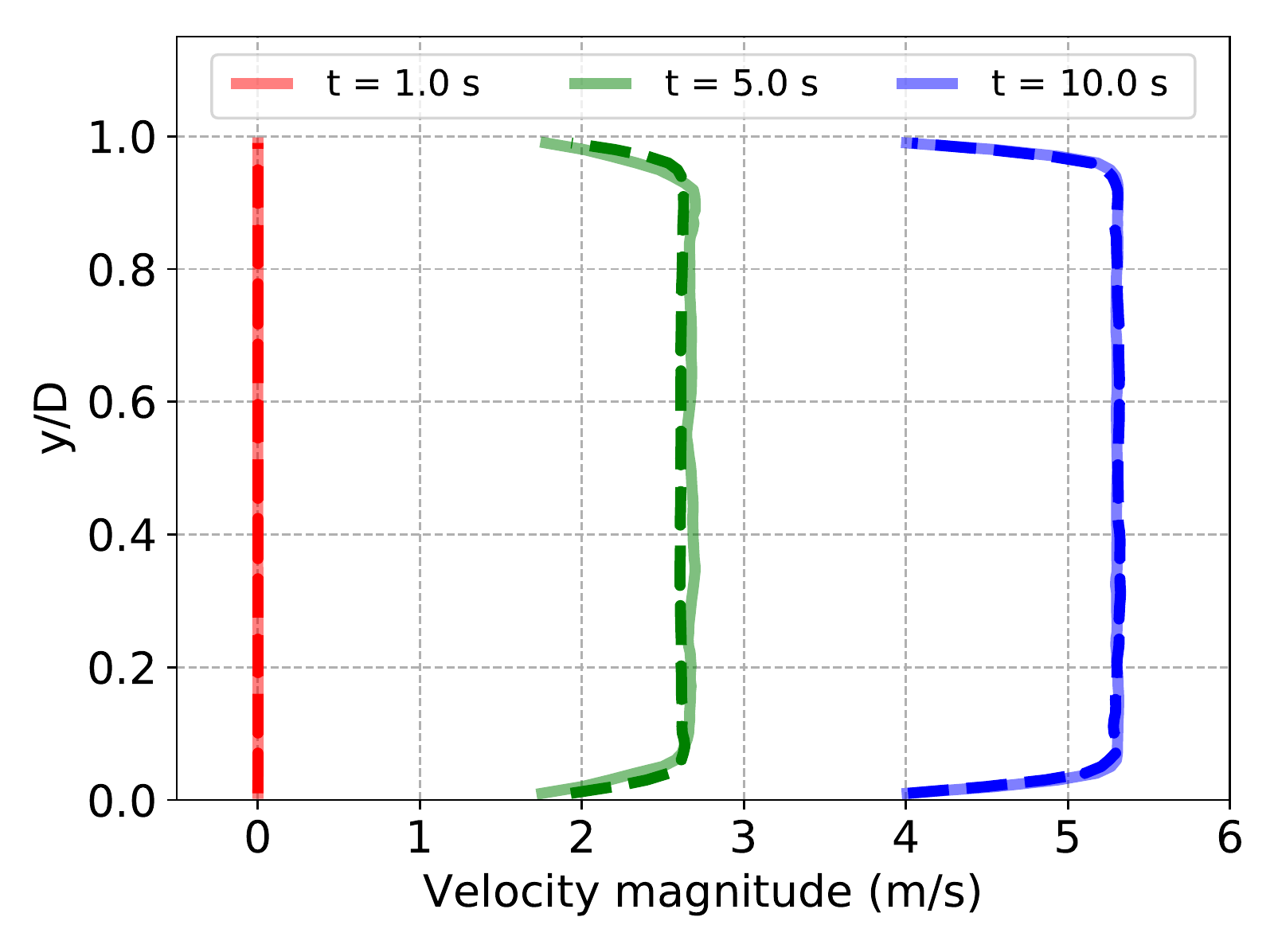}
\caption{Profiles of the velocity magnitude at $L_{CFD}/D$ = 0.5 downstream of the inlet of the CFD domain at $t$ = 1.0, 5.0 and 10.0 s of simulation time of the closed pipe flow test with a maximum pump rotor rotational speed of 100 rad/s. Solid lines: RELAP5/CFD; Dashed lines: RELAP5/ROM.}
\label{fig:profiles_closed_vel}
\end{figure}

\newpage
Figure~\ref{fig:profiles_closed_vel} shows the profiles of the velocity magnitude in the CFD domain obtained with the RELAP5/CFD and RELAP5/ROM coupled models at $L_{CFD}/D$ = 0.5 at $t$ = 1.0, 5.0 and 10.0 s of simulation time. The RELAP5/ROM velocity profiles are in good agreement with the RELAP5/CFD profiles. At $t$ = 5.0 s the profiles are slightly deviating from each other, while they are almost fully overlapping at $t$ = 10.0 s. This corresponds to Figure~\ref{fig:res_closed_rel_v}, which shows that the velocity relative error decreases towards the final simulation time.
\newpage

\begin{figure}[h!]

\centering
\includegraphics[width=0.7\linewidth]{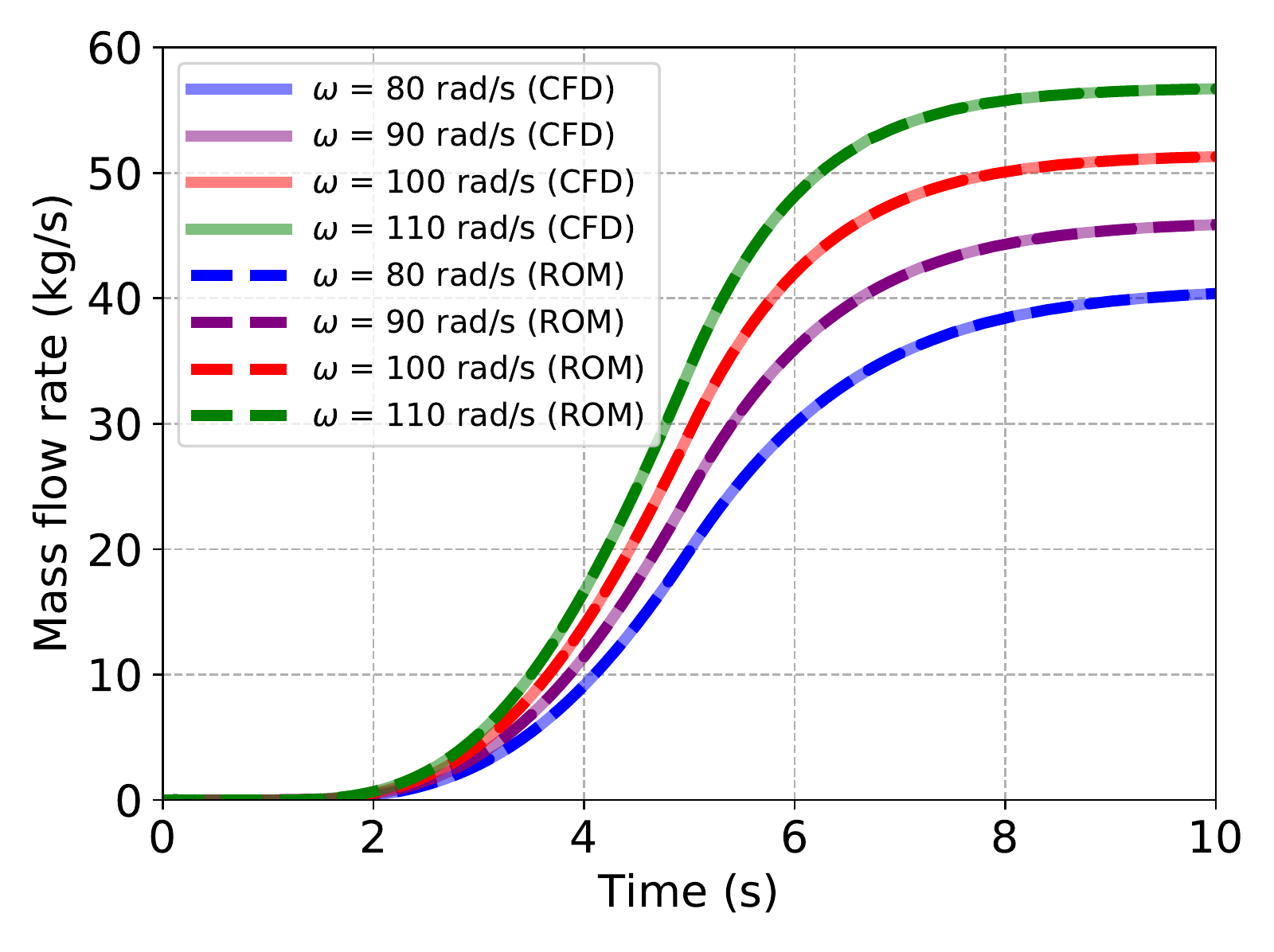}
\caption{Time evolution of the mass flow rate through the pipe at interface 2 for the closed pipe flow test for different maximum pump rotor rotational speeds. $N_r = 10$ modes are used for the ROM.}
\label{fig:res_closed_para_phi}
\end{figure}

\begin{figure}[h!]
\centering
\includegraphics[width=0.7\linewidth]{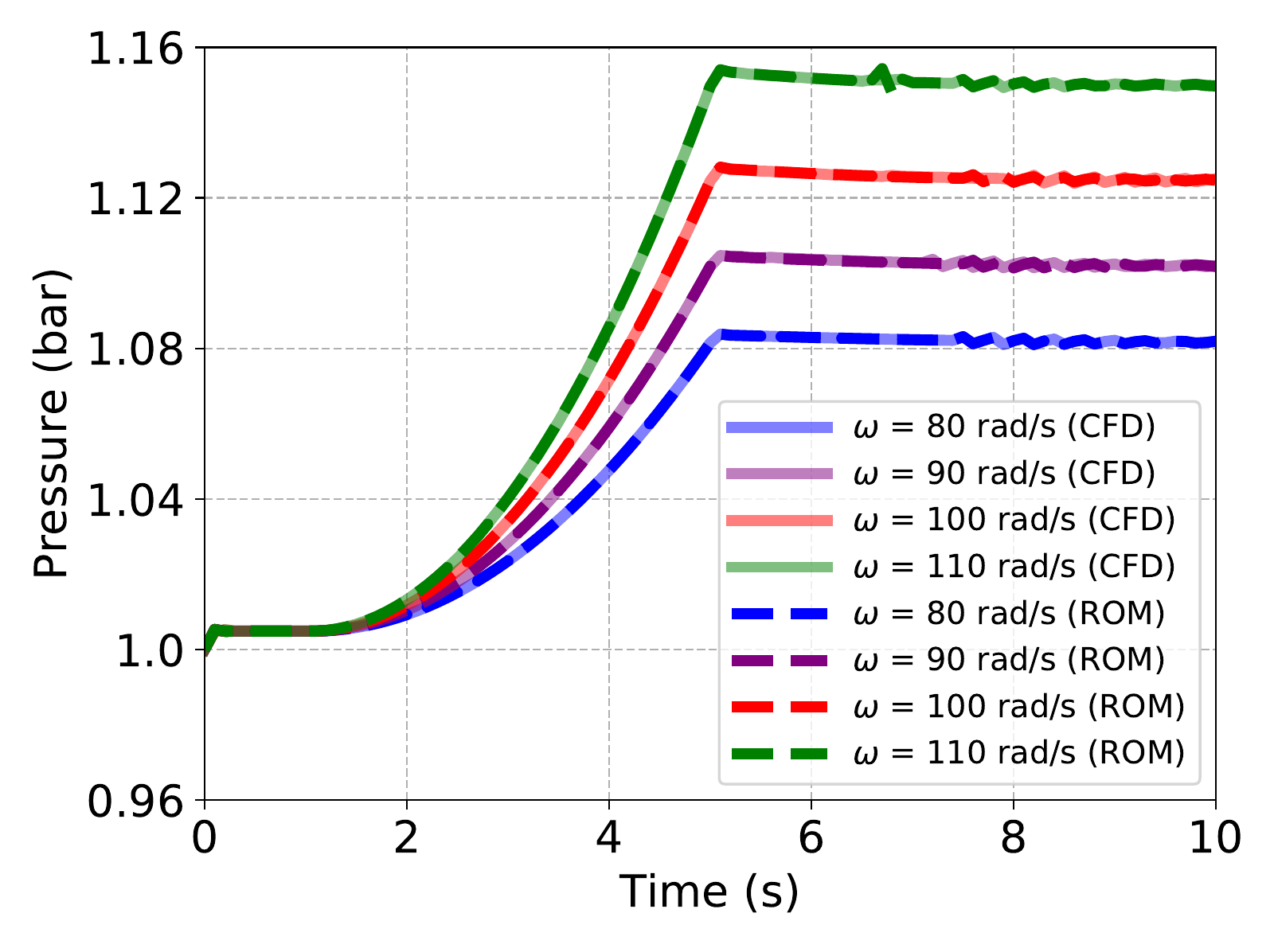}

\caption{Time evolution of the pressure at interface 2 for the closed pipe flow test for different maximum pump rotor rotational speeds. $N_r = 10$ modes are used for the ROM.}
\label{fig:res_closed_para_p}
\end{figure}

The closed loop test case is also used to test the coupled RELAP5/ROM model, of which the reduced basis is constructed with snapshots obtained for $\omega$ = 100 rad/s, on a parametric problem. Figures~\ref{fig:res_closed_para_phi} and~\ref{fig:res_closed_para_p} show the mass flow rate and the pressure evaluated at interface $\Gamma_2$ for the maximum pump rotor rotational speed of 80, 90, 100 and 110 rad/s, respectively. The results are compared with those obtained with the RELAP5/CFD model for the same rotational speeds.


\newpage

Finally, one coupled RELAP5/CFD simulation for the closed pipe flow test takes 3.2 $\cdot$ \num{e3} seconds, while a coupled RELAP5/ROM simulation takes 6.5 $\cdot$ \num{e2} seconds to complete. Thus, the obtained speed-up is about 3.5 times.

\subsection{Convergence history}
\begin{figure}[h!]
\centering

\includegraphics[width=0.7\linewidth]{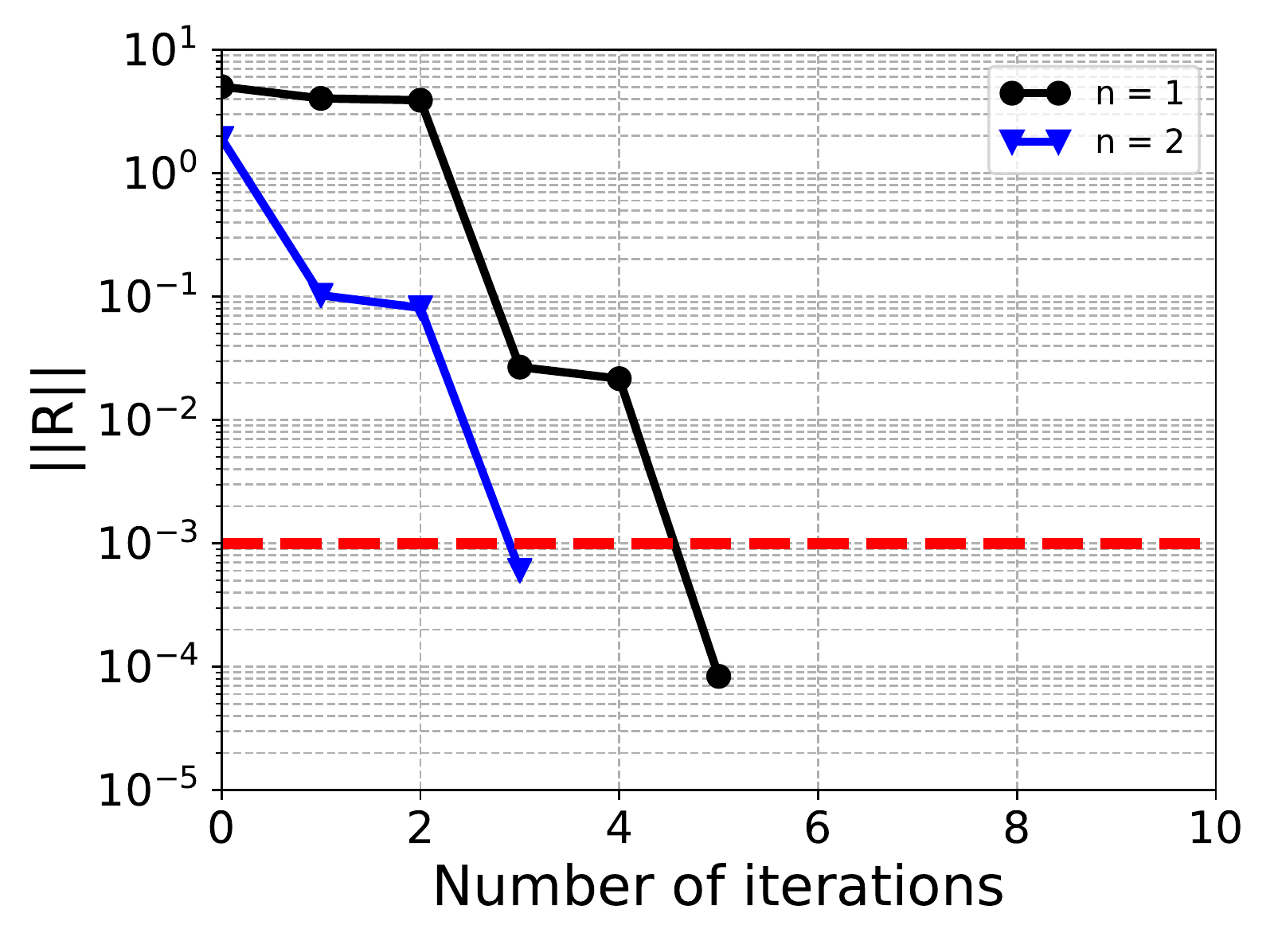}
\caption{Interface convergence history for the first two time steps of the open pipe flow test case.}
\label{fig:res_closed_time_step_open}
\end{figure}

\begin{figure}[h!]
\centering
\includegraphics[width=.7\linewidth]{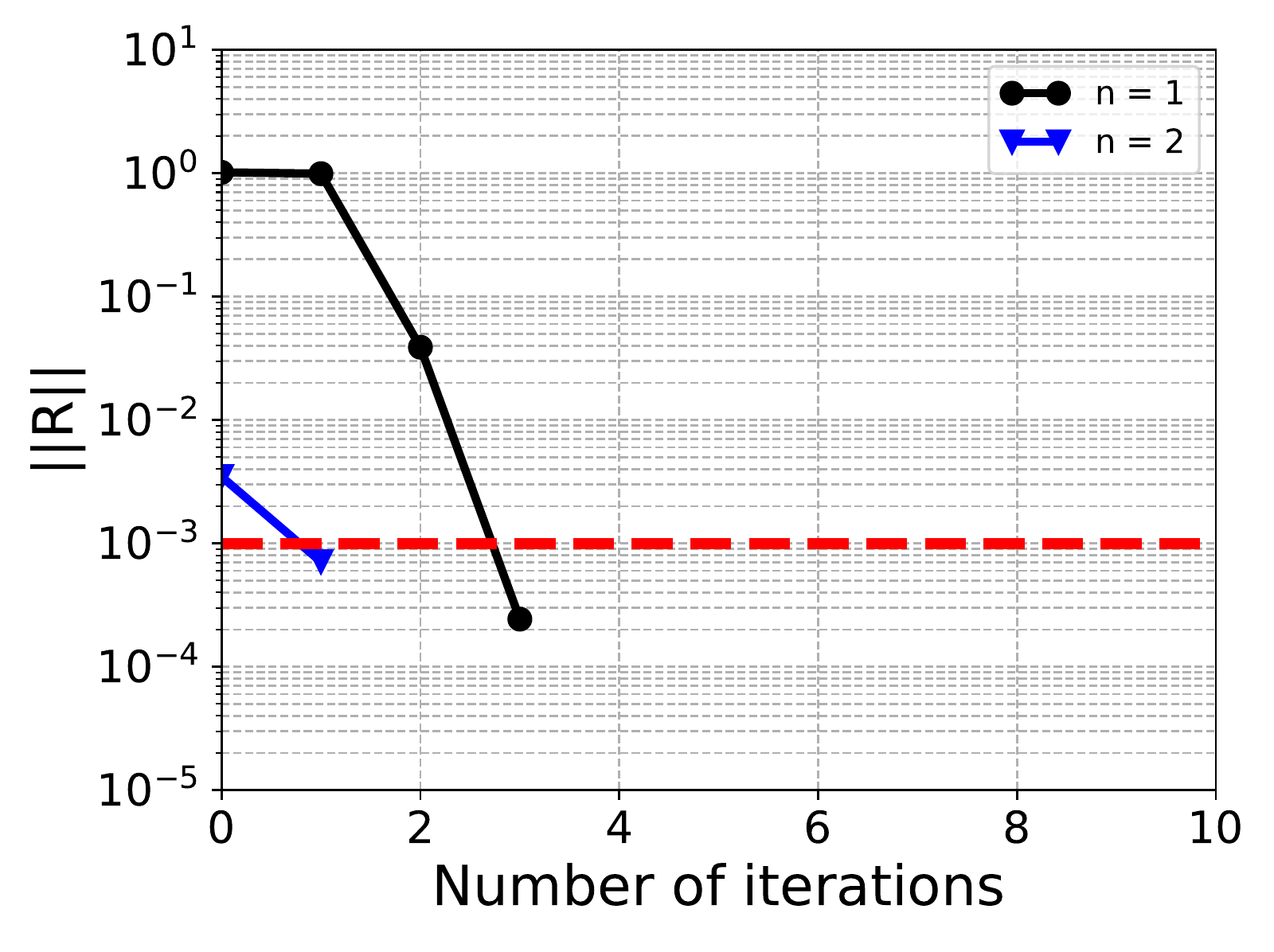}

\caption{Interface convergence history for the first two time steps of the closed pipe flow test case.}
\label{fig:res_closed_time_step_closed}
\end{figure}

The performance of the implicit coupling algorithm is analyzed by checking the number of iterations needed to reach a residual norm below \num{e-3}. Figures~\ref{fig:res_closed_time_step_open} and~\ref{fig:res_closed_time_step_closed} show the convergence history for the first two coupling time steps for the open pipe test and for the closed pipe test, respectively. In case of the open pipe flow test case, the first time step requires five iterations for the residual to drop below the threshold and the second time step requires three iterations. For the closed loop test case, the number of iterations is reduced to three for the first time step and only one for the second time step. Therefore, the convergence rate is higher for the closed pipe flow test than the open pipe flow test. Previous results of the relative error (Figures~\ref{fig:res_open_rel_v} and~\ref{fig:res_closed_rel_v}) also showed that this test case is numerically more stable than the open pipe flow test.
For both cases the Jacobian is not available in the first iteration of the first time step, which explains the difference in number of iterations for the first and second time step. In case of the closed pipe flow test, the Jacobian computed in the first time step is also used in the second time step as Equation~\ref{eq:Rcriteria} is satisfied~\cite{toti2018coupled}. Therefore, the convergence criterion is met with only one additional iteration.

\newpage
\section{Discussion}\label{sec:discussion}
The RELAP5/CFD models exhibit numerical perturbations in the form of oscillations throughout the coupled simulations. As concluded in previous works on coupled CFD codes with 1D system codes~\cite{toti2018coupled,grunloh2016novel}, these perturbations are caused by the overestimation of the mass flow rate at the coupling interfaces in the first few time steps of the simulations.
A change in pressure drop will immediately affect the whole solution domain as the fluid density does not change with pressure unlike in compressible fluids. These small oscillations are also present in the results of the RELAP5/ROM models. Nevertheless, the ROMs remain stable, i.e.\ the ROM results do not blow up over time, even though small perturbations in reduced order models can lead to unphysical results~\cite{Lassila}. Toti et al.~\cite{toti2016development} concluded that reducing the time step of the coupled simulations affects how rapidly numerical oscillations are damped during the transients. However, the maximum pressure oscillation amplitude is independent from the time step.

For all test cases, only one coupled RELAP5/CFD simulation for a single parameter value has been performed for collecting the snapshots. The ROMs are capable of predicting solutions for new parameter values and for long time integration as (part of) the flow features for these new cases are similar to the snapshots. However, to be able to analyze and to improve the accuracy of the ROM for a larger parameter range, snapshots from multiple simulations for different parameter values are required to construct the reduced basis spaces. When several sets of snapshots are required for the POD, one can optimize the POD procedure by using a nested POD approach~\cite{georgaka2018parametric}.

In this work, a uniform velocity profile is implemented at the inlet of the CFD domain. Toti~\cite{toti2018coupled} et al. considered in their work for open and closed pipe flow test that the error introduced by the uniform profile is small as the velocity distribution across the section of the pipe is fairly flat in the case of fully developed turbulent pipe flow. However, as one of the main purposes of developing coupled models that accurately quantify the flow fields in specific parts of the computational domain, it is better to take into account the curvature of the velocity profiles in case of complex flow problems.

The relative velocity error is about two orders lower at final simulation time in the case of the closed pipe flow test compared to the open pipe flow test. The closed loop is less prone to numerical instabilities due to the absence of interruptions in the STH domain. Therefore, perturbations are transported throughout the whole STH domain within a single coupling iteration, while several coupling iterations are required in the case of the open pipe test as the upstream and downstream parts of STH sub-domain are hydraulically decoupled. Because of that, the convergence rate is generally higher for the closed pipe than for the open pipe test. Reduced order models are sensitive to these numerical instabilities what explains the difference in relative errors for these two test cases. As system analysis of nuclear installations, like MYRRHA, are dealing mainly with closed cooling loops, it is advantageous that numerical instabilities are less prone for closed loop systems. 

Furthermore, the results for pressure fields calculated in the CFD sub-domain are about two order worse than those for the velocity fields. This has been observed in previous work of Stabile et al.~\cite{Stabile2017CAF}. Rather than using the PPE methods, a supremizer enrichment technique can be used to improve the pressure fields.

\section{Conclusions and outlook}\label{sec:conclusion}
The best-estimate system thermal-hydraulic code RELAP5 is coupled with the finite volume CFD solver OpenFOAM and its reduced order model. The codes are coupled implicitly by a partitioned domain decomposition coupling algorithm in which the hydraulics variables are exchanged between the sub-domains at the coupling boundary interfaces. 

The ROM is constructed with a finite volume based POD-Galerkin projection method. The average velocity determined at the single junction of the STH sub-domain at coupling interface 1 is imposed at the inlet boundary of the reduced order model with a boundary control method, namely a penalty method.

Academic tests are carried out on open and closed pipe flow configurations. The coupled RELAP5/ROM models accurately predict the time evolution of the mass flow rate and pressure results of the coupled RELAP5/CFD models at one of the coupling interfaces. Also for new conditions, the RELAP5/ROM models are capable of reproducing the behavior of the RELAP5/CFD models at the coupling interface. In addition, the RELAP5/ROM model for reversed flow in a closed loop flow test case performs well for long time integration. The effect of the number of POD modes on the accuracy of the RELAP5/ROM results is not investigated in this work.

The pressure results exhibit numerical perturbations in the form of oscillations throughout the coupled simulations, both with the CFD solver and the reduced order model. Nevertheless, they do not lead to a blow up of the RELAP5/ROM results, even though small perturbations in reduced order models can, generally, lead to unphysical results.

Finally, the coupled RELAP5/ROM simulations are about 3 to 5 times faster than the coupled RELAP5/CFD simulations performed on a single Intel\textsuperscript{\tiny\textregistered} Xeon\textsuperscript{\tiny\textregistered} core. Therefore, it is shown that the computational cost of coupled STH/CFD models can be reduced by replacing the CFD solver by a reduced order model. Furthermore, the coupled RELAP5/ROM model can be used to study a number of different conditions at a lower computational cost compared to the coupled RELAP5/CFD model.

In future work, the methodology needs to be extended to reduced order models for turbulent buoyancy driven flows, for which the discretized momentum and energy equations are coupled in a two-way manner~\cite{vergari2020reduced,starMC2019}. In addition, the models need to be adjusted for low-Prandtl number flows~\cite{star2020pod}, such as LBE. The coupled models also need to be validated against experimental results, such as a loss of flow due to a pump trip transient in the TALL-3D experimental facility~\cite{toti2017improved}. Moreover, the coupling could be extended to parallel computing to speed up the simulations.

\section*{Acknowledgment}\label{sec:ack}
The authors would like to thank the ITHACA-FV developers and contributors for their input and insightful discussions on the development of reduced order models. In particular, Giovanni Stabile, Saddam Hijazi and Matteo Zancanaro from SISSA mathLab, Umberto Morelli from ITMATI and Sokratia Georgaka from Imperial College London. 

\section*{CRediT author statement}
\noindent \textbf{S.K. Star}: Conceptualization, Methodology, Visualization, Writing - original draft. \textbf{G. Spina}: Methodology, Software,  Formal analysis, Investigation, Visualization, Writing - review \& editing. \textbf{F. Belloni}: Writing - review \& editing, Project administration. \textbf{J. Degroote}: Writing - review \& editing, Project administration, Funding acquisition.

\bibliographystyle{ieeetr}
\bibliography{mybibfile}  

\end{document}